\documentclass[11pt,a4paper]{article}
\usepackage{ajtex}

\usepackage{tikz}
\usetikzlibrary{decorations.markings,decorations.pathmorphing,arrows.meta}

\tikzset{mom/.style={line width=0.5mm}}
\tikzset{aux/.style={decorate,decoration={snake,segment length=1.5mm,
      amplitude=0.4mm}}}
\tikzset{coil/.style={decorate,decoration={coil, aspect=0.7, amplitude = 0.5mm,
      segment length = 1mm}}}

\tikzset{left/.style={arrows={Stealth[scale length=0.5, scale width=1.5][sep=2pt]-}}}
\tikzset{right/.style={arrows={-[sep=-2pt]Stealth[scale length=0.5, scale
      width=1.5]}}}

\def\be{\begin{equation}} 
\def\ee{\end{equation}}
\def\bes{\begin{equation*}} 
\def\ees{\end{equation*}}
\def\bea{\begin{eqnarray}}
\def\eea{\end{eqnarray}}
\def\bal{\begin{align}} 
\def\eal{\end{align}}
\def\lb{\label}

\def\mc{\mathcal}

\def\nn{\nonumber}
\def\df{\mathrm{d}}

\def\v[#1]{\boldsymbol{#1}}
\def\w[#1]{\widehat{#1}}
\def\vs[#1,#2]{\boldsymbol{{#1}_{#2}}}
\def\mes[#1]{d^{3}{#1}}
\def\del{\partial}
\def\<{\langle}
\def\>{\rangle}
\def\vecs[#1,#2]{\boldsymbol{{#1}_{#2}}}

\newcommand{\BD}[1]{{\dot \bar{1}}}

\def\l{\lambda}
\def\L{\Lambda}
\def\m{\mu}
\def\n{\nu}
\def\N{\nabla}

\def\vp{\varphi}
\def\s{\sigma}

\newcommand{\nc}{\newcommand}
\nc{\al}{\alpha}
\nc{\ga}{\gamma}
\nc{\de}{\delta}
\nc{\ep}{\epsilon}
\nc{\ze}{\zeta}
\nc{\et}{\eta}

\nc{\ka}{\kappa}
\nc{\la}{\lambda}
\nc{\rh}{\rho}
\nc{\si}{\sigma}
\nc{\ta}{\tau}
\nc{\up}{\upsilon}
\nc{\ph}{\phi}
\nc{\ch}{\chi}
\nc{\ps}{\psi}
\nc{\om}{\omega}
\nc{\Ga}{\Gamma}
\nc{\De}{\Delta}
\nc{\La}{\Lambda}
\nc{\Si}{\Sigma}
\nc{\Up}{\Upsilon}
\nc{\Ph}{\Phi}
\nc{\Ps}{\Psi}
\nc{\Om}{\Omega}
\nc{\ptl}{\partial}
\nc{\bi}{\bibitem}

\def\lb{\left(}
\def\rb{\right)}
\def\lB{\left[}
\def\rB{\right]}
\def\nn{\nonumber}
\def\dow{\partial}

\title{Hydrodynamic effective field theory and the analyticity of hydrostatic correlators}

\author[a]{Akash Jain}\email{ajain@uvic.ca} 
\author[a]{Pavel Kovtun}\email{pkovtun@uvic.ca}
\author[a]{Adam Ritz} \email{aritz@uvic.ca}
\author[\,b,\,a]{Ashish Shukla}\email{ashukla@perimeterinstitute.ca}

\affiliation[a]{Department of Physics \& Astronomy, University of Victoria, 3800
  Finnerty Road, Victoria, \\ \, British Columbia V8P 5C2, Canada}
\affiliation[b]{Perimeter Institute for Theoretical Physics, 31 Caroline Street North, Waterloo, Ontario\\ \, N2L 2Y5, Canada}

\abstract{We study one-loop corrections to retarded and symmetric hydrostatic
  correlation functions within the Schwinger-Keldysh effective field theory
  framework for relativistic hydrodynamics, focusing on charge diffusion. We
  first consider the simplified setup with only diffusive charge density
  fluctuations, and then augment it with momentum fluctuations in a model where
  the sound modes can be ignored. We show that the loop corrections, which
  generically induce non-analyticities and long-range effects at finite
  frequency, non-trivially preserve analyticity of retarded correlation
  functions in spatial momentum due to the KMS constraint, as a manifestation of
  thermal screening. For the purposes of this analysis, we develop an
  interacting field theory for diffusive hydrodynamics, seen as a limit of
  relativistic hydrodynamics in the absence of temperature and longitudinal
  velocity fluctuations.}

\begin{document}

\maketitle


\section{Introduction}
\label{intro}

Hydrodynamics provides an effective description of near-equilibrium thermal
states, by focusing on the conserved charges and their transport
properties. However, despite its generality, hydrodynamics is incomplete as a
true {\it effective field theory} (EFT), as it accounts only for dissipation but
not the stochastic fluctuations that are present even in equilibrium. Over the
past few years, there has been significant progress in our understanding of how
to combine the symmetries and the relevant fluctuation degrees of freedom into a
complete hydrodynamic EFT, building off the microscopic Schwinger-Keldysh (SK)
formalism \cite{Grozdanov:2013dba, Kovtun:2014hpa, Harder:2015nxa, Crossley:2015evo, Haehl:2015uoc,
  Jensen:2017kzi, Jain:2020vgc} (see \cite{Glorioso:2018wxw} for a pedagogical
review). Proper consideration of stochastic fluctuations has already led to
qualitatively new results in the non-equilibrium (non-hydrostatic) regime. For
instance, it was recently discovered that new ``stochastic transport
coefficients'' emerge at higher derivative orders due to stochastic interactions
that have no analogue in ``classical'' hydrodynamics, but nonetheless affect the
low-energy behaviour of hydrodynamic correlation
functions~\cite{Jain:2020fsm}. EFT tools have also been recently used to revisit
long-time tails due to density fluctuations at one-loop
order~\cite{Chen-Lin:2018kfl}.

Despite these qualitative departures from classical hydrodynamics, one expects
that the EFT results will still boil down to their well-known hydrostatic form
as we dial back to equilibrium. In particular, it has been appreciated that
careful consideration of the equilibrium limit provides a powerful set of
consistency conditions on the hydrodynamic framework~\cite{Banerjee:2012iz,
  Jensen:2012jh}. These conditions restrict the hydrodynamic constitutive
relations in a manner that was understood previously to be due only to a local
formulation of the second law of thermodynamics. Focusing on relativistic
hydrodynamics with a conserved U(1) charge, in a nutshell, one demands that the
equilibrium values of the conserved energy-momentum tensor $T^{\mu\nu}$ and
charge current $J^\mu$ can be derived from a hydrostatic generating functional,
${\cal W} = T^{-1}_0 \int d^d x \sqrt{-g}\,{\cal F}(g_{\mu\nu},A_\mu)$,
expressed in terms of the background metric $g_{\mu\nu}$ and gauge field sources
$A_\mu$. Here $T_0$ is the constant temperature of the global thermal state and
$d$ is the number of spatial dimensions. To wit,
\begin{equation}
  \de {\cal W}[g_{\mu\nu},A_\mu] = T^{-1}_0 \int d^d x \sqrt{-g} \,
  \left( \frac12 \, T_{\text{eq}}^{\mu\nu} \de g_{\mu\nu}
    + J_{\text{eq}}^\mu \de A_\mu \right).
\end{equation}
The associated conservation equations $\N_\m T^{\m\n} = F^{\n\l} J_\l$ and
$\N_\m J^\m = 0$ follow directly from the diffeomorphism and gauge invariance of
$\mc{W}$. An important ingredient in implementing the ensuing constraints on the
hydrodynamic constitutive relations is the finiteness of the static correlation
length. This feature is expected for generic thermal systems that are not at
second order critical points and do not have spontaneously broken symmetries,
and implies that retarded hydrostatic (time-independent) correlation functions
should fall-off exponentially at large distances. Equivalently, the Fourier
space zero-frequency retarded correlation functions must be analytic in an
expansion in small spatial momentum. The generating functional thus admits a
derivative expansion, and one can obtain the hydrodynamic constitutive relations
in the static limit order-by-order in this expansion.

Moving away from thermal equilibrium, the full EFT formalism of hydrodynamics
should capture the physics of small fluctuations about the hydrostatic
equilibrium state. Thus, hydrodynamic correlation functions should reduce to
those derived from the equilibrium generating functional on integrating out all
the fluctuation modes. Of particular interest is the requirement of analyticity
at small spatial momenta noted above, that follows directly from thermal
screening and the finite static correlation length. In the presence of
stochastic interactions, it is known that hydrodynamic correlation functions are
generically non-analytic out of equilibrium due to the presence of long-time
tails~\cite{PhysRevLett.36.867}. The requirement that these long-distance or
infrared non-analyticities must drop out at all loop orders as we specialize to
hydrostatic configurations, is quite non-trivial and poses a rigid consistency
check for the EFT framework. This question is particularly important because the
tree-level propagators for the fluctuation degrees of freedom in these EFTs lack
any intrinsic scale that could define the finite spatial correlation length. In
fact, the lack of such a scale is precisely what leads to the physics of
long-time tails out of equilibrium.

A crucial ingredient in the EFT framework is the discrete Kubo-Martin-Schwinger
(KMS) symmetry~\cite{Crossley:2015evo}. It is motivated from its namesake KMS
condition in thermal field theory~\cite{doi:10.1143/JPSJ.12.570,
  PhysRev.115.1342}, and ensures that the non-linear fluctuation-dissipation
theorem~\cite{Wang:1998wg} is satisfied by the EFT framework. While the KMS
condition captures the fact that the system in question is fluctuating around
the thermal state, by itself it does not guarantee the finiteness of the spatial
correlation length. However, combined with the structure of hydrodynamic EFTs,
we find that the KMS symmetry does, in fact, conspire in a nontrivial way to
ensure the analyticity of hydrostatic correlators. The goal of this paper is to
illustrate how this analyticity emerges explicitly at one-loop order in
relativistic hydrodynamics for density-density two-point functions.

For simplicity, we will work in a kind of ``incompressible limit'', where sound
modes are frozen and the theory only has longitudinal charge diffusion and
transverse shear diffusion modes in the low-energy spectrum (see \cite{PhysRevA.16.732} for an early discussion along this direction). Since sound modes
are typically ``higher energy'' compared to diffusive modes, we expect that the
simplified model faithfully captures the low-energy behaviour of the
EFT. Existing notions of ``incompressibility'' in the literature, however, are
only consistent with linearised hydrodynamics. As we include interactions, due
care has to be taken while defining this limit so as to not violate the
underlying structure of the EFT, in particular the KMS condition. To reaffirm
the role of KMS symmetry in ensuring the analyticity of hydrostatic correlators,
for the majority of this paper we will lift the KMS symmetry by untying the
dissipative and stochastic fluctuation parameters in the EFT. The simplified
model that we develop in this paper, which we dub \emph{diffusive
  hydrodynamics}, will also be helpful more generally to probe stochastic
signatures in hydrodynamics in the presence of interacting momentum modes. In
the remainder of this introductory section, we will elaborate further on the
analyticity features of interest, and also summarize our results.

\subsection{Thermal correlators and summary of results}
\label{summary}

Recall that for a generic set of bosonic Hermitian operators $\mathcal{O}_I$,
the retarded and symmetric correlators are defined respectively as
\begin{align}
  G^{\mathrm R}_{IJ}(t-t',{\bf x} - {\bf x}')
  &\equiv i\theta(t-t') \langle [\mc{O}_I(t,{\bf x}), \mc{O}_J(t',{\bf x}')]\rangle,
    \nonumber\\
  G^{\mathrm S}_{IJ}(t-t',{\bf x} - {\bf x}')
  &\equiv \frac{1}{2} \langle \{\mc{O}_I(t,{\bf x}), \mc{O}_J(t',{\bf x}')\}\rangle.
\end{align}
The thermal expectation values are taken in the grand canonical ensemble with
the Hamiltonian $H' = H - \mu Q$, where $Q$ is a conserved charge. These correlators are not independent. For
example, working in Fourier space and specializing to the hydrodynamic
regime $\omega/T_0 \ll 1$, one can derive the relation
\begin{equation}
  G^{\mathrm S}_{IJ}(\om,{\v[k]})
  = \frac{2T_0}{\omega}\, {\rm Im} \,G^{\rm R}_{IJ}(\om,{\v[k]}).
  \label{fdt}
\end{equation}
This is an incarnation of the fluctuation-dissipation theorem, which ties
together statistical fluctuations in a thermal system to dissipation, and
follows from the KMS periodicity condition in Euclidean time for equilibrium
states.

For generic systems in thermal equilibrium, at least away from second order
critical points and in the absence of any spontaneously broken symmetries,
screening implies a finite static or spatial correlation length. It follows that
retarded (or Euclidean) correlation functions fall off exponentially at large
distances, or equivalently the Fourier space retarded correlators at
zero-frequency are analytic at low spatial momentum $\v[k]$, i.e.
\begin{equation}
  G^{\rm R}_{IJ}(\om=0,{\v[k]}) \,\,
  \mbox{is analytic as ${\v[k]}\rightarrow 0$}.
  \label{analytic}
\end{equation}
We refer to this condition as {\it spatial analyticity}. Importantly, it allows
for a well-defined derivative expansion, used previously to develop a
hydrostatic generating functional for hydrodynamics~\cite{Banerjee:2012iz,
  Jensen:2012jh}.  In the hydrodynamic regime, where $\om/T_0 \ll 1$,
$|{\v[k]}|/T_0 \ll 1$ and we consider perturbations about an equilibrium state,
the combination of \cref{fdt,analytic} implies a nontrivial structure for the
retarded correlator.

In this paper, we will focus on the correlations of a conserved charge
$n = J^0$, which is the time-component of a conserved $U(1)$ current $J^\m$.  We
can investigate the above features by looking at the symmetric and retarded
correlators for $n$. At the tree level in the hydrodynamic EFT (i.e. in the
linear response theory), they are known to take the form~\cite{Kovtun:2012rj}
\begin{align}
  G_{nn}^{{\rm R},\text{tree}}(\om, {\v[k]})
  &= \frac{i\chi D {\v[k]}^2}{\om + iD {\v[k]}^2}
    \quad\implies\quad
    G_{nn}^{{\rm R},\text{tree}}(\om\rightarrow 0, {\v[k]})
    = \ch + \frac{i\chi \om}{D {\v[k]}^2} + {\cal O}(\om^2), \nonumber\\
  G_{nn}^{{\rm S},\text{tree}}(\om, {\v[k]})
  &= \frac{2T\tilde\sigma {\v[k]}^2}{\om^2 + (D{\v[k]}^2)^2}
    \quad\implies\quad
    G_{nn}^{{\rm S},\text{tree}}(\om\rightarrow 0, {\v[k]})
    = \frac{2\tilde\sigma T}{D^2 {\v[k]}^2} + {\cal O}(\om^2).
    \label{eq:tree-correlators}
\end{align}
In terms of the chemical potential $\mu$ and conductivity $\sigma$, the
coefficients appearing here are the static charge susceptibility
$\chi = \dow n/\dow \mu|_T$ and the diffusion constant $D = \s/\chi$. The
coefficient $\tilde\sigma$ controls the strength of stochastic interactions. In
thermal equilibrium, the KMS condition sets $\tilde\sigma = \sigma$, leading to
these tree-level correlators obeying the fluctuation-dissipation theorem in
\cref{fdt}. We note that the symmetric correlator at $\om=0$, which is nothing
but the tree-level static propagator of $n$, has a pole at ${\v[k]}=0$. This is
consistent with our discussion above because symmetric correlators are not
restricted by any analyticity requirement. Such poles are, of course, a generic
feature of gapless degrees of freedom. The relevant point here is that despite
the gapless nature of these tree-level correlations, the retarded correlator is
indeed analytic at zero frequency in accordance with \cref{analytic}.

While the KMS condition does not play any role at tree-level in ensuring the
analyticity of the retarded correlator, it becomes crucial as we start to
include loop corrections. In the hydrodynamic EFT, upon accounting for density
and transverse-momentum fluctuations within diffusive hydrodynamics, and
specializing to $d=3$ spatial dimensions, we find that the correlators in
\cref{eq:tree-correlators} admit one-loop corrections that behave at $\omega=0$
as
\begin{align}
  {G}^{\rm R}_{nn} (\om=0, {\v[k]})
  &= \chi \Big(Z_\chi + c^{\rm R}_1|\v[k]| + c^{\rm R}_2|\v[k]|^3
    + {\cal O}({\v[k]}^4) \Big), \\
  G^{\rm S}_{nn}(\om=0,{\v[k]})
  &= \frac{2T\chi}{D\v[k]^2} \Big(
    Z_\Lambda + c_1^{\rm S}|\v[k]|
    + c_2^{\rm S} |\v[k]|^3
    + {\cal O}({\v[k]}^4) \Big).
    \label{eq:correction-summary}
\end{align}
Here $Z_{\chi,\Lambda}$ are UV-sensitive renormalization factors with polynomial
dependence on the momentum-cutoff of the EFT. These are expected in the
low-energy effective field theory due to the irrelevant nature of hydrodynamic
interactions. We shall not compute these factors in detail in this paper; a
thorough discussion restricted to the scalar diffusion model can be found
in~\cite{Chen-Lin:2018kfl}.

The remaining coefficients in \cref{eq:correction-summary} are interesting
physically, because they quantify the infrared and non-analytic corrections to
the hydrodynamic correlation functions. These can be explicitly computed in
terms of three hydrodynamic parameters: pressure $p(T,\mu)$, shear viscosity
$\eta(T,\mu)$, and conductivity $\sigma(T,\mu)$, and two stochastic parameters
$\tilde\eta(T,\mu)$ and $\tilde\sigma(T,\mu)$. The enthalpy density $w(T,\mu)$
and charge density $n(T,\mu)$ are defined using the usual thermodynamic
relations $\df p = (w-\mu n)/T\, \df T + n\df\mu$. The bulk viscosity
$\zeta(T,\mu)$ and the associated stochastic parameter $\tilde\zeta(T,\mu)$ do
not appear in the diffusive limit of hydrodynamics. Using the charge
susceptibility $\chi = \dow n/\dow\mu|_T$ and the diffusion constant
$D = \sigma/\chi$ from above, and further defining the shear diffusion constant
$\gamma_\eta = \eta/w$ and the cross-susceptibility
$\chi_\epsilon = \dow w/\dow\mu|_T$, we find
\begin{align}
  c_1^{\rm R}
  &= \frac{T}{32\pi D^2w}
    \lb \frac{\tilde\eta}{\eta} - \frac{\tilde\sigma}{\sigma} \rb\lB
    \frac{2\sqrt{\gamma_\eta D} (D-\gamma_\eta)}{(\gamma_\eta +D)^2}
    + \arccos\lb\frac{\gamma_\eta-D}{\gamma_\eta +D}
    \rb\rB, \nn\\
  c_2^{\rm R}
  &= \frac{-T}{32\pi D\chi}
    \lB \frac{\dow D}{\dow \mu} \frac{\dow(\tilde\sigma/\sigma)}{\dow\mu}
    + \frac{1}{w}\frac{\dow(\tilde\eta/\eta)}{\dow\mu}
    \lb (3\pi-2)\chi_\epsilon D + (7\pi-20/3) n \gamma_\eta \rb \rB, \nn\\
  c_1^{\rm S}
  &= \frac{T}{32\pi D^2 w}
    \frac{\tilde\sigma}{\sigma} \frac{\tilde\eta}{\eta}
    \lB
    \frac{2\sqrt{\gamma_\eta D}(D-\gamma_\eta)}{(\gamma_\eta +D)^2}
    + \arccos\lb\frac{\gamma_\eta-D}{\gamma_\eta +D}\rb \rB, \nn\\
  c_2^{\rm S}
  &= \frac{-T}{32\pi D\chi} \frac{\tilde\sigma}{\sigma}
    \lB \frac{\tilde\sigma}{\sigma} \frac{1}{D} \bfrac{\dow D}{\dow\mu}^2
    + \frac{1}{w^2\gamma_\eta}\frac{\tilde\eta}{\eta}
    \lb \pi \chi^2_\epsilon D^2 + (5\pi-4) n \chi_\epsilon D \gamma_\eta
    + 2(3\pi - 10/3) n^2 \gamma_\eta^2
    \rb \rB \nn\\
  &\qquad
    - \frac{T}{32\pi w^2\chi D\gamma_\eta} \frac{\tilde\eta}{\eta}
    \lb\frac{\tilde\eta}{\eta} - \frac{\tilde\sigma}{\sigma}\rb 
    \lb \pi\chi^2_\epsilon D^2 - 2(3\pi - 10/3) n^2 \gamma_\eta^2 \rb \nn\\
  &\qquad
    + \frac{T}{16\pi w\chi D} \frac{\tilde\sigma}{\sigma}
    \frac{\dow(\tilde\eta/\eta)}{\dow\mu}
    \lb (3\pi - 2) \chi_\epsilon D
    + (7\pi - 20/3) n \gamma_\eta \rb.
\end{align}
All the coefficients here are understood to be evaluated on the equilibrium
configuration. The explicit functional form of these coefficients is not very
important. The important point is that both the retarded and symmetric
correlators are generically non-analytic in the static limit. However, upon
imposing the KMS condition, which in this case requires setting
$\tilde\sigma = \sigma$ and $\tilde\eta = \eta$, we find
\begin{align}
  c_1^{\rm R}
  &\overset{\text{KMS}}{=} 0, \qquad
    c_2^{\rm R}
    \overset{\text{KMS}}{=} 0, \nn\\
  c_1^{\rm S}
  &\overset{\text{KMS}}{=} \frac{T}{32\pi D^2 w}
    \lB
    \frac{2\sqrt{\gamma_\eta D}(D-\gamma_\eta)}{(\gamma_\eta +D)^2}
    + \arccos\lb\frac{\gamma_\eta-D}{\gamma_\eta +D}\rb \rB, \nn\\
  c_2^{\rm S}
  &\overset{\text{KMS}}{=}
    \frac{-T}{32\pi D\chi} 
    \lB \frac{1}{D} \bfrac{\dow D}{\dow\mu}^2
    + \frac{1}{w^2\gamma_\eta}
    \lb \pi \chi^2_\epsilon D^2 + (5\pi-4) n \chi_\epsilon D \gamma_\eta
    + 2(3\pi - 10/3) n^2 \gamma_\eta^2
    \rb \rB.
\end{align} 
We see that in the presence of the KMS condition, which ensures that the
correlators satisfy the fluctuation-dissipation theorem, the non-analytic pieces
in the retarded correlator drop out in the hydrostatic limit, reaffirming the
consistency of hydrostatic equilibrium in the EFT framework. On the other hand,
the symmetric correlator in the hydrostatic limit remains non-analytic. Full
analysis of the finite-$\omega$ behaviour of these correlation functions and the
ensuing long-time tails will appear in a companion paper~\cite{other-paper}.

The discussion for arbitrary spatial dimensions follows in a similar manner. In
$d\neq 1,2$ spatial dimensions,\footnote{Note that in $d=1$ spatial dimension,
  the leading correction to the retarded correlator goes as $1/|\v[k]|$ and is
  relevant, signaling the fact that the hydrodynamic description does not
  apply~\cite{2002PhRvL..89t0601N}.}  the corrections inside the brackets in
\cref{eq:correction-summary} typically behave as
$|\v[k]|^{d-2}, |\v[k]|^{d},\ldots$, while in $d=2$ spatial dimensions, they
behave as $\log(\v[k]^2), |\v[k]|^{2}\log(\v[k]^2),\ldots$. While we will not
derive the explicit form of the associated coefficients in general, we will show
in the course of this paper that the one-loop non-analytic corrections to the
retarded function drop out in arbitrary dimensions due to the KMS symmetry,
maintaining consistency with the existence of thermal equilibrium.

The paper is organized as follows. To sketch out the road-map and develop a
qualitative understanding of the results, we will first ignore the momentum
modes altogether in \cref{sec:diffusiveEFT}, and start with a simplified EFT
model of a single diffusive conserved charge as proposed
in~\cite{Chen-Lin:2018kfl, Jain:2020fsm}. However, unlike the original
references, we will lift the KMS symmetry from the model and compute the
one-loop corrections to the symmetric and retarded correlation functions
directly using the EFT generating functional. Note that we cannot deduce the two
correlators using one-another in the absence of KMS symmetry, because the
fluctuation-dissipation theorem in \cref{fdt} does not apply. We will then set up
the theory of diffusive hydrodynamics in \cref{sec:diff-hydro}, and repeat the
computation of one-loop correlation functions in the presence of transverse
momentum fluctuations. We will finish with a discussion in
\cref{sec:discussion}. In \cref{details}, we provide further details of the
derivation of the EFT for diffusive hydrodynamics and the explicit one-loop
computations.


\section{Density correlation functions in simple diffusion}
\label{sec:diffusiveEFT}

In this section, we investigate the issue of spatial analyticity of hydrostatic
correlators in a simple EFT model with a single diffusive charge, which has all
the relevant qualitative features of the full problem. The generalization to
full hydrodynamics will be presented in the next section.

\subsection{Simple stochastic diffusion}

Thermal fluctuations are conventionally introduced into the hydrodynamic setup
by including stochastic noise sources in the constitutive relations with
short-range Gaussian correlations. One way to realize this effective theory is
to add a fluctuation field for each conserved hydrodynamic variable in the
theory and require that the ensuing effective action, when inserted inside a
path integral, reproduces the full set of real-time correlation functions in
thermal equilibrium~\cite{LandLif, PhysRevA.8.423, landau1980course}
(see~\cite{Kovtun:2012rj} for a review). Focusing on a single conserved charge
$n(\mu)$ with the chemical potential $\mu$, the respective classical
conservation equation at one derivative order is given by
\begin{equation}
  \partial_t n(\mu) + \dow_iJ^i = 0, \qquad
  J^i = - \sigma(\mu)\, \delta^{ij}(\partial_j\mu - F_{jt}),
\end{equation}
where $J^i$ is the charge flux expressed in terms of $\mu$ and its derivatives.
In the following we will often use the ``covariant'' current $J^\mu$ with
$J^0 = n$. Here $F_{it} = \dow_i A_t - \dow_t A_i$ is the electric field
associated with the background gauge field $A_\mu$ coupled to $J^\mu$. The
charge conductivity $\sigma(\mu)$ is an arbitrary function of $\mu$, required to
be non-negative by the second law of thermodynamics. The equilibrium state is
given by $\mu = \mu_0$ being a constant.

In the EFT language, the chemical potential $\mu$ is related to the fundamental
``chemical shift'' field $\Lambda_\beta$ as $\mu = T_0\Lambda_\beta + A_{rt}$,
where $A_{r\mu}$ is one of the Schwinger-Keldysh background gauge fields along
with $A_{a\mu}$.\footnote{\label{foot:phir}In the EFT language
  of~\cite{Crossley:2015evo}, the true fundamental degree of freedom is the
  phase field $\varphi_r$, which is related to $\Lambda_\beta$ above via the
  definition $\Lambda_\beta T_0 = \mu_0 + \dow_t\varphi_r$. For an ordinary
  fluid phase where the U(1) symmetry is not spontaneously broken, the system
  respects a spatial chemical shift symmetry:
  $\varphi_r(x) \to \varphi_r(x) - \lambda(\v[x])$, and all the shift-invariant
  information in $\varphi_r$ is contained in $\Lambda_\beta$.}  To account for
stochastic fluctuations, we add an auxiliary field $\varphi_a$ to partner the
hydrodynamic field $\Lambda_\beta$. The effective action $S$ of the theory is
fixed so that it reproduces all the two-point response functions at tree-level
as predicted by classical linearized hydrodynamics. Introducing the effective
Lagrangian density through $S = \int d^{d+1}x \, \mc{L}$, we
have~\cite{Crossley:2015evo, Jain:2020fsm}
\begin{align}
  {\cal L}
  &=  J^\mu \lb \dow_\mu \varphi_a + A_{a\mu} \rb
    + iT_0\tilde{\sigma}(\mu) \lb \dow_i \varphi_a + A_{ai} \rb
    \lb \dow^i \varphi_a + A_{a}^i \rb.
    \label{L1}
\end{align}
Here $T_0$ is the constant temperature of the thermal equilibrium state. We have
introduced an arbitrary coefficient $\tilde\sigma(\mu)$ that characterizes the
strength of stochastic fluctuations. This is the most general Lagrangian
consistent with the Schwinger-Keldysh EFT framework for a single diffusive
charge at one-derivative order. However, for this effective Lagrangian to
correctly reproduce the hydrodynamic response functions, we need to impose the
constraint from the fluctuation-dissipation theorem. This equates the strength
of stochastic fluctuations $\tilde\sigma$ to the dissipative transport
coefficient $\sigma$ (see~\cite{Kovtun:2012rj} for more details),
\be
 {\rm KMS:} \quad \tilde\sigma = \sigma.
\ee
 Assuming the
underlying microscopic theory to be CPT-invariant, in the Schwinger-Keldysh (SK)
effective field theory framework of~\cite{Crossley:2015evo}, this follows as a
consequence of the dynamical KMS symmetry of the effective action
\begin{gather}
  \mu_0 \to -\mu_0, \qquad
  \Lambda_\beta(x) \to -\Lambda_\beta(-x), \qquad
  \varphi_a(x) \to \varphi_a(-x) + i\Lambda_\beta(-x), \nn\\
  A_{r\mu}(x) \to - A_{r\mu}(-x), \qquad
  A_{a\mu}(x) \to - A_{a\mu}(-x) - iT_0^{-1}(\dow_t A_{r\mu})(-x).
  \label{eq:KMS-diff}
\end{gather}
In addition, it implies the charge conjugation properties for the coefficients
$n(-\mu) = -n(\mu)$, $\sigma(-\mu)=\sigma(\mu)$, and
$\tilde\sigma(-\mu)=\tilde\sigma(\mu)$.  In the following, we find that the KMS
condition is critical for ensuring the analyticity of retarded correlators in
spatial momentum. Hence, to keep track of the non-analytic behaviour explicitly,
we will perform the majority of forthcoming manipulations assuming
$\tilde\sigma$ and $\sigma$ to be independent.

\subsection{Linear expansion and interactions}

To be able to use this theory in a loop expansion, we need to expand the
Lagrangian \eqref{L1} order-by-order in fluctuations about the equilibrium state
$\Lambda_\beta = \mu_0/T_0$, $\varphi_a=0$; see~\cite{Jain:2020fsm} for
details. It is practically easier to work with the fluctuations in the density
$\delta n = n(T_0\Lambda_\beta) - n(\mu_0)$ rather than
$\Lambda_\beta$. Ignoring the background fields at first, and ignoring certain
total-derivative terms, we find the ``free'' Gaussian part of the effective
Lagrangian to be
\begin{equation}
  \mathcal{L}_{\text{free}}
  = - \varphi_a \left( \dow_t\delta n - D \N^2 \delta n \right)
  + iT \tilde{\s} \dow^i\varphi_a\dow_i\varphi_a,
  \label{Ldiff-free}
\end{equation}
where $D = \sigma/\chi$ is the diffusion constant, with $\chi = n'(\mu)$ being
the charge susceptibility, and $\N^2 \equiv \del^i \del_i$. We have dropped the subscript ``0'' for clarity and the coefficients are understood to be evaluated at equilibrium, i.e.
$\mu = \mu_0$. This leads to the tree-level propagators
\begin{align}
  \langle \delta n(p)\varphi_a(-p)\rangle_0
  &= \frac{1}{\omega + iD\v[k]^2},
  && \raisebox{-3mm}{\begin{tikzpicture}[thick]
      \draw [right] (0mm,0)  -- (10mm,0);
      \draw [aux] (10mm,0)  -- (20mm,0);
      \node at (10mm,-3mm) {$p$};
    \end{tikzpicture}} \nn\\
  \langle \varphi_a(p)\delta n(-p)\rangle_0
  &= \frac{-1}{\omega - iD\v[k]^2},
  && \raisebox{-3mm}{\begin{tikzpicture}[thick]
      \draw [aux,right] (0mm,0)  -- (10mm,0);
      \draw (10mm,0)  -- (20mm,0);
      \node at (10mm,-3mm) {$p$};
    \end{tikzpicture}} \nn\\
  \langle \delta n(p)\delta n(-p)\rangle_0
  &= \frac{2T\tilde\sigma\v[k]^2}{\omega^2 + D^2\v[k]^4}
    = \frac{iT\chi\tilde\sigma/\sigma}{\omega+iD\v[k]^2}
    - \frac{iT\chi\tilde\sigma/\sigma}{\omega-iD\v[k]^2},
  && \raisebox{-3mm}{\begin{tikzpicture}[thick]
      \draw [right] (0mm,0)  -- (10mm,0);
      \draw (10mm,0)  -- (20mm,0);
      \node at (10mm,-3mm) {$p$};
    \end{tikzpicture}} \nn\\
  \langle \varphi_a(p)\varphi_a(-p)\rangle_0
  &= 0.
     \label{eq:diff-propagators}
\end{align}
Here $p=(\omega,\v[k])$ collectively denotes frequency and momentum.  We have
denoted $\delta n$ by solid and $\varphi_a$ by wavy lines. Note that the
$\langle\delta n \varphi_a\rangle$ propagator is purely retarded,
$\langle\varphi_a\delta n\rangle$ propagator is purely advanced, while
$\langle\delta n\delta n\rangle$ propagator can be expressed as a sum of
retarded and advanced pieces. This observation generalizes to full
hydrodynamics, and will be important in the explicit loop calculations in the
following discussion.

The free diffusive action \eqref{Ldiff-free} is well-known in the literature
(see e.g.~\cite{Kovtun:2012rj}), however, it does not account for
interactions. To this end, we can expand \cref{L1} to the next order in
$\delta n$ and $\varphi_a$ to obtain\footnote{See \cite{Kovtun:2014nsa, Chao:2020kcf} for discussions along similar lines, involving the dependence of transport parameters on the fluctuating fields.}
\begin{align}
  {\cal L}^{\text{3pt}}_{\text{int}}
  &= \half \l\, \delta n^2\, \N^2\varphi_a 
    + i\chi T \tilde\lambda\,\delta n\, \dow_i\varphi_a \dow^i\varphi_a,
    \label{L1int}
\end{align}
where the two coupling constants $\lambda$ and $\tilde\lambda$ are defined via
\begin{equation}
  \lambda = \frac1\chi \frac{\dow D}{\dow \mu}, \qquad
  \tilde\lambda = \frac{1}{\chi^2} \frac{\dow\tilde\sigma}{\dow \mu}.
\end{equation}
We can work-out the momentum-space Feynman rules for the associated vertices
\begin{align}
  &\raisebox{-9mm}{\begin{tikzpicture}[thick]
      \draw [aux,left] (-10mm,0)  -- (0,0);
      \node at (-7mm,-3mm) {$p$};
      \draw [right] (0,0)  -- (7mm,7mm);
      \node at (3mm,7mm) {$p'$};
      \draw [right] (0,0)  -- (7mm,-7mm);
      \node at (3mm,-7mm) {$p''$};
    \end{tikzpicture}}
    \qquad \frac{-i}{2}\lambda\, \v[k]^2, 
  &&\raisebox{-9mm}{\begin{tikzpicture}[thick]
      \draw [left] (-10mm,0)  -- (0,0);
      \node at (-7mm,-3mm) {$p$};
      \draw [aux,right] (0,0)  -- (7mm,7mm);
      \node at (3mm,7mm) {$p'$};
      \draw [aux,right] (0,0)  -- (7mm,-7mm);
      \node at (3mm,-7mm) {$p''$};
    \end{tikzpicture}}
     \qquad T\chi\tilde\lambda\, \v[k]'\cdot \v[k]''.
     \label{eq:vertices-diff}
\end{align}
The same procedure can be iterated to arbitrarily high orders in fluctuations,
depending on the sensitivity required. This form of the effective action, albeit
for energy diffusion instead of charge diffusion, was derived recently
in~\cite{Chen-Lin:2018kfl}.

To be able to compute the correlations functions using the non-equilibrium generating
functional \cite{Crossley:2015evo}, we also need the structure of the background field couplings in the
EFT. Since, in this work, we are only interested in density correlators, we only
turn on the $A_{rt}$, $A_{at}$ components of the gauge fields. We directly
borrow the results from~\cite{Jain:2020fsm}, leading to
\begin{align}
  \mathcal{L}_{\text{source}}^{\text{2pt}}
  &= \delta n\, A_{at}
    + \chi \dow_t\varphi_a A_{rt}
    + \chi A_{rt}A_{at}, \nn\\
  \mathcal{L}_{\text{source}}^{\text{3pt}}
  &=  \frac{\chi'}{\chi} \delta n\, A_{rt}A_{at}
    + \half\chi' \dow_t\varphi_a A_{rt}^2
    + \half\chi' A_{rt}^2A_{at} \nn\\
  &\qquad
    + \lb \frac{\chi'}{\chi} \delta n\, \dow_t\varphi_a
    - \frac{1}{\chi} \frac{\dow\sigma}{\dow\mu}
    \partial^i \delta n \dow_i\varphi_a
    + iT \frac{\dow\tilde\sigma}{\dow\mu} \dow_i \varphi_a \dow^i \varphi_a
    \rb A_{rt},
    \label{L1source}
\end{align}
where $\chi' = \dow\chi/\dow\mu$.  Denoting $A_{rt}$ by a dotted and $iA_{at}$
by a dashed line, we can represent these diagrammatically as
\begin{align}
  &\raisebox{-4mm}{\begin{tikzpicture}[thick]
      \draw [right] (0mm,0)  -- (0.1mm,0);
      \draw [dashed] (0mm,0)  -- (10mm,0);
      \draw [right] (10mm,0)  -- (20mm,0);
      \node at (10mm,-3mm) {$p$};
    \end{tikzpicture}}
    \qquad 1,
  && \raisebox{-4mm}{\begin{tikzpicture}[thick]
      \draw [right] (0mm,0)  -- (0.1mm,0);
      \draw [aux] (0mm,0)  -- (10mm,0);
      \draw [dotted,right] (10mm,0)  -- (20mm,0);
      \node at (10mm,-3mm) {$p$};
    \end{tikzpicture}}
       \qquad -\omega\chi, \nn\\
  &\raisebox{-4mm}{\begin{tikzpicture}[thick]
      \draw [right] (0mm,0)  -- (0.1mm,0);
      \draw [dashed] (0mm,0)  -- (10mm,0);
      \draw [dotted,right] (10mm,0)  -- (20mm,0);
      \node at (10mm,-3mm) {$p$};
    \end{tikzpicture}}
    \qquad \chi,
  &&\raisebox{-9mm}{\begin{tikzpicture}[thick]
      \draw [right,dotted] (0,0)  -- (9mm,0);
      \node at (7mm,-3mm) {$p$};
      \draw [dashed,left] (-7mm,7mm) -- (0,0);
      \node at (-3mm,8mm) {$p'$};
      \draw [left] (-7mm,-7mm) -- (0,0);
      \node at (-3mm,-7mm) {$p''$};
    \end{tikzpicture}}\qquad
     \frac{\chi'}{\chi},  \nn\\
  &\raisebox{-9mm}{\begin{tikzpicture}[thick]
      \draw [right,dotted] (0,0)  -- (9mm,0);
      \node at (7mm,-3mm) {$p$};
      \draw [dotted,left] (-7mm,7mm) -- (0,0);
      \node at (-3mm,8mm) {$p'$};
      \draw [aux,left] (-7mm,-7mm) -- (0,0);
      \node at (-3mm,-7mm) {$p''$};
    \end{tikzpicture}}\qquad
    \half \chi' \omega'',
  &&\raisebox{-9mm}{\begin{tikzpicture}[thick]
      \draw [right,dotted] (0,0)  -- (9mm,0);
      \node at (7mm,-3mm) {$p$};
      \draw [dotted,left] (-7mm,7mm) -- (0,0);
      \node at (-3mm,8mm) {$p'$};
      \draw [dashed,left] (-7mm,-7mm) -- (0,0);
      \node at (-3mm,-7mm) {$p''$};
    \end{tikzpicture}}\qquad
     \half \chi', \nn\\
    &\raisebox{-9mm}{\begin{tikzpicture}[thick]
      \draw [right,dotted] (0,0)  -- (9mm,0);
      \node at (7mm,-3mm) {$p$};
      \draw [aux,left] (-7mm,7mm) -- (0,0);
      \node at (-3mm,8mm) {$p'$};
      \draw [left] (-7mm,-7mm) -- (0,0);
      \node at (-3mm,-7mm) {$p''$};
    \end{tikzpicture}}\qquad
    \frac{\chi'}{\chi}\omega'
    + \frac{i}{\chi}\frac{\dow\sigma}{\dow\mu} \v[k]'\cdot\v[k]'',
  &&\raisebox{-9mm}{\begin{tikzpicture}[thick]
      \draw [right,dotted] (0,0)  -- (9mm,0);
      \node at (7mm,-3mm) {$p$};
      \draw [aux,left] (-7mm,7mm) -- (0,0);
      \node at (-3mm,8mm) {$p'$};
      \draw [aux,left] (-7mm,-7mm) -- (0,0);
      \node at (-3mm,-7mm) {$p''$};
    \end{tikzpicture}}\qquad
     T\frac{\dow\tilde\sigma}{\dow\mu} \v[k]'\cdot\v[k]''.
\end{align}
The first two of these are the usual linear couplings between operators and
sources in quantum field theories, while the remaining are non-linear
couplings. Frequency and momenta going away from the vertex are taken to be
positive.

Let us briefly comment on the dimension counting scheme that we implement in the
following discussion. For diffusive processes, it is typically argued that
$\dow_t \sim \dow^i \dow_i$. Following e.g.~\cite{Kaplan:2005es}, we can demand
that $[\dow_i]=1$ and require that the couplings in spatial kinetic terms in
\cref{Ldiff-free} are dimensionless. If we take $[\dow_t]=2$ to ensure
consistent scaling, and thus $[{\cal L}]=d+2$, we find
$[\varphi_a]=[\delta n]=d/2$, and it follows that
$[\lambda]=[\tilde\lambda]=-d/2$ and thus both interactions in \cref{L1int} are
irrelevant. We therefore expect that loops will induce power-like sensitivity to
the UV cutoff. To count loops ($L$), we can use the relation $L=1+(N/2-1)V-E/2$
for $N^{th}$ order vertices in diagrams with $E$ external lines. In the case of
interest below, $E=2$, and so at one-loop order, $V=2/(N-2)$. Thus, one-loop
corrections require diagrams with 2 cubic vertices. In principle, diagrams with
a single quartic vertex can also contribute at this order, however the
contributions from such diagrams to the two-point functions are purely UV-cutoff
dependent and only renormalize the classical hydrodynamic
parameters~\cite{Chen-Lin:2018kfl, Jain:2020fsm}.

\subsection{One-loop corrections from density fluctuations}

Defining the Schwinger-Keldysh generating functional as
\cite{Crossley:2015evo}\footnote{\label{foot:jacobian}Technically, the EFT path
  integral should be performed over the phase field $\varphi_r$, with the field
  space volume divided by the associated spatial chemical shift symmetry; see
  \cref{foot:phir}. Passing over to $\delta n$ or $\Lambda_\beta$ might
  generically result in additional Jacobian factors that we shall ignore for
  now.}
\begin{equation}
  \exp W[A_{r\mu},A_{a\mu}]
  = \int \mathcal{D}\delta n\mathcal{D}\varphi_a
  \exp\lb i \int \df^{d+1} x\,\mathcal{L} \rb,
\end{equation} 
we can compute the retarded and symmetric correlators of $n$ via
\begin{align}
  G^{\text{R}}_{nn}(p)
  &= \frac{-i\delta^2W}{\delta A_{at}(p)\delta A_{rt}(-p)}
    = \chi - \omega \chi 
    \langle \delta n(p) \varphi_a(-p) \rangle
    + (\text{source coupling diagrams}), \nn\\
  G^{\text{S}}_{nn}(p)
  &= \frac{-\delta^2W}{\delta A_{at}(p)\delta A_{at}(-p)}
    =  \langle \delta n(p) \delta n(-p) \rangle.
    \label{eq:variational_formulas}
\end{align}
Using the propagators in \cref{eq:diff-propagators}, it is trivial to check that
we reproduce the tree-level results for the two correlation functions given in
\cref{eq:tree-correlators}.

We will now compute the stochastic loop corrections to the classical tree-level
propagators in eq.\ \eqref{eq:tree-correlators}.  This amounts to integrating
out the stochastic noise field $\vp_a$ at one-loop order using the interaction
vertices defined above in eqs.\ \eqref{L1int} and \eqref{L1source}.  Let us
start with the retarded function. At one-loop order, we have two diagrams
contributing to the $\langle \delta n\varphi_a\rangle$ propagator
(see~\cite{Jain:2020fsm} for details)
\begin{subequations}
  \begin{equation}
    \begin{aligned}
      & \begin{tikzpicture}[thick] \draw [right] (0mm,0) -- (5mm,0); \draw [aux]
        (5mm,0) -- (10mm,0); \draw [right] (10mm,0) arc (-180:-90:6mm); \draw
        (22mm,0) arc (0:-90:6mm); \draw [right] (10mm,0) arc (180:90:6mm); \draw
        [aux] (22mm,0) arc (0:90:6mm); \draw [right] (22mm,0) -- (27mm,0); \draw
        [aux] (27mm,0) -- (32mm,0);
      \end{tikzpicture}\qquad
      && \begin{tikzpicture}[thick] \draw [right] (0mm,0) -- (5mm,0); \draw
        [aux] (5mm,0) -- (10mm,0); \draw [right] (10mm,0) arc (-180:-90:6mm);
        \draw [aux] (22mm,0) arc (0:-90:6mm); \draw [right] (10mm,0) arc
        (180:90:6mm); \draw [aux] (22mm,0) arc (0:90:6mm); \draw [right]
        (22mm,0) -- (27mm,0); \draw [aux] (27mm,0) -- (32mm,0);
      \end{tikzpicture}
    \end{aligned}
    \label{eq:diff-ret-diagrams-prop}
  \end{equation}
  In addition, we have two diagrams involving non-linear background source
  couplings
  \begin{equation}
    \begin{aligned}
      & \begin{tikzpicture}[thick] \draw [right] (0mm,0) -- (5mm,0); \draw [aux]
        (5mm,0) -- (10mm,0); \draw [right] (10mm,0) arc (-180:-90:6mm); \draw
        (22mm,0) arc (0:-90:6mm); \draw [right] (10mm,0) arc (180:90:6mm); \draw
        [aux] (22mm,0) arc (0:90:6mm); \draw [right, dotted] (22mm,0) --
        (28mm,0);
      \end{tikzpicture}\qquad\quad
      && \begin{tikzpicture}[thick] \draw [right] (0mm,0) -- (5mm,0); \draw
        [aux] (5mm,0) -- (10mm,0); \draw [right] (10mm,0) arc (-180:-90:6mm);
        \draw [aux] (22mm,0) arc (0:-90:6mm); \draw [right] (10mm,0) arc
        (180:90:6mm); \draw [aux] (22mm,0) arc (0:90:6mm); \draw [right, dotted]
        (22mm,0) -- (28mm,0);
      \end{tikzpicture}
    \end{aligned}
    \label{eq:diff-ret-diagrams-back}
  \end{equation}
  \label{eq:diff-ret-diagrams}%
\end{subequations}
We can compute these diagrams to find the one-loop correction to the retarded
density correlator; see \cref{app:loop_details} for calculational details. We find the one-loop correction to the retarded correlator to be
\begin{equation}
  G^{{\rm R},\text{1-loop}}_{nn}(p)
  = \frac{iT\lambda \v[k]^2 }{(\omega + iD\v[k]^2)^2}
  \lb  \frac{\tilde\sigma}{\sigma} \omega \chi^2  \lambda
  - (\omega + iD\v[k]^2) \sigma \frac{\dow(\tilde\sigma/\sigma)}{\dow\mu}
  \rb \mathcal{I}_1(p),
  \label{eq:diff-R-1loop}
\end{equation}
where $p=(\omega,\v[k])$ and $\mathcal{I}_1(p)$ is the result of the loop
momentum integral
\begin{align}
  \mathcal{I}_1(p)
  = \frac{\v[k]^2}{2} \int^\Lambda \frac{\df^{d}\v[k]'}{(2\pi)^d}
  \frac{1}{i\omega - D(\v[k]'^2 + (\v[k]-\v[k]')^2)} 
  &\overset{d=3}{=}
    \frac{\v[k]^2}{32 \pi D} \sqrt{\v[k]^2-\frac{2 i \omega}{D} }
    + \mathcal{O}(\Lambda), \nn\\
  &\overset{d=2}{=}
    \frac{\v[k]^2}{16\pi D} \ln \lb \v[k]^2-\frac{2 i \omega}{D} \rb
    + \mathcal{O}(\ln \Lambda).
    \label{eq:I1}
\end{align}
The value of this integral depends on the UV-cutoff $\Lambda$ and the number of
spatial dimensions $d$. Above, we have explicitly computed it for $d=2,3$ for
reference. We see that, generically, the one-loop corrections at finite
frequency exhibit IR-sensitive non-analytic dependence on $\omega$, reflecting
the physics of long-time tails, which invalidates the hydrodynamic derivative
expansion at a given order \cite{Kovtun:2011np}. However, if we restrict to the hydrostatic ($\om=0$)
limit, we have argued that thermal screening should render the correlator
analytic. Setting $\omega=0$, the one-loop correction reduces to
\begin{equation}
  G^{{\rm R},\text{1-loop}}_{nn}(0,\v[k])
  = - T\chi \lambda
  \, \frac{\dow(\tilde\sigma/\sigma)}{\dow\mu}
  \, \mathcal{I}_1(0,\v[k])
  \overset{\text{KMS}}{=} 0,
\end{equation}
which vanishes trivially upon imposing the KMS condition $\tilde\sigma = \sigma$,
irrespective of the dimensionality. Hence, we see that the KMS condition is
crucial to reproduce the analyticity of retarded correlators in the hydrostatic limit.

We can repeat the same procedure for the symmetric correlator. In this case, we
do not have any background coupling diagrams at one-loop order. We only need
to consider the seven diagrams correcting the $\langle \delta n\delta n \rangle$
propagator
\begin{subequations}
  \begin{gather}
    \begin{gathered}
      \begin{tikzpicture}[thick]
        \draw [right] (0mm,0) -- (5mm,0); \draw [aux] (5mm,0) -- (10mm,0); \draw
        [right] (10mm,0) arc (-180:-90:6mm); \draw (22mm,0) arc (0:-90:6mm);
        \draw [right] (10mm,0) arc (180:90:6mm); \draw [aux] (22mm,0) arc
        (0:90:6mm); \draw [right] (22mm,0) -- (27mm,0); \draw (27mm,0) --
        (32mm,0);
      \end{tikzpicture}\qquad
      \begin{tikzpicture}[thick]
        \draw [right] (0mm,0) -- (5mm,0); \draw [aux] (5mm,0) -- (10mm,0); \draw
        [right] (10mm,0) arc (-180:-90:6mm); \draw [aux] (22mm,0) arc
        (0:-90:6mm); \draw [right] (10mm,0) arc (180:90:6mm); \draw [aux]
        (22mm,0) arc (0:90:6mm); \draw [right] (22mm,0) -- (27mm,0); \draw
        (27mm,0) -- (32mm,0);
      \end{tikzpicture}
      \label{eq:diff-sym-diagrams-1}
    \end{gathered} \\
    \begin{gathered}
      \begin{tikzpicture}[thick]
        \draw [right] (0mm,0) -- (5mm,0); \draw (5mm,0) -- (10mm,0); \draw
        [right] (10mm,0) arc (-180:-90:6mm); \draw (22mm,0) arc (0:-90:6mm);
        \draw [right, aux] (10mm,0) arc (180:90:6mm); \draw (22mm,0) arc
        (0:90:6mm); \draw [right, aux] (22mm,0) -- (27mm,0); \draw (27mm,0) --
        (32mm,0);
      \end{tikzpicture}\qquad
      \begin{tikzpicture}[thick]
        \draw [right] (0mm,0) -- (5mm,0); \draw (5mm,0) -- (10mm,0); \draw
        [aux,right] (10mm,0) arc (-180:-90:6mm); \draw (22mm,0) arc (0:-90:6mm);
        \draw [aux,right] (10mm,0) arc (180:90:6mm); \draw (22mm,0) arc
        (0:90:6mm); \draw [aux,right] (22mm,0) -- (27mm,0); \draw (27mm,0) --
        (32mm,0);
      \end{tikzpicture}
      \label{eq:diff-sym-diagrams-2}
    \end{gathered} \\
    \begin{gathered}
      \begin{tikzpicture}[thick]
        \draw [right] (0mm,0) -- (5mm,0); \draw [aux] (5mm,0) -- (10mm,0); \draw
        [right] (10mm,0) arc (-180:-90:6mm); \draw (22mm,0) arc (0:-90:6mm);
        \draw [right] (10mm,0) arc (180:90:6mm); \draw (22mm,0) arc (0:90:6mm);
        \draw [right, aux] (22mm,0) -- (27mm,0); \draw (27mm,0) -- (32mm,0);
      \end{tikzpicture}\qquad
      \begin{tikzpicture}[thick]
        \draw [right] (0mm,0) -- (5mm,0); \draw [aux] (5mm,0) -- (10mm,0); \draw
        [right] (10mm,0) arc (-180:-90:6mm); \draw (22mm,0) arc (0:-90:6mm);
        \draw [right] (10mm,0) arc (180:90:6mm); \draw [aux] (22mm,0) arc
        (0:90:6mm); \draw [right, aux] (22mm,0) -- (27mm,0); \draw (27mm,0) --
        (32mm,0);
      \end{tikzpicture}\qquad
      \begin{tikzpicture}[thick]
        \draw [right] (0mm,0) -- (5mm,0); \draw [aux] (5mm,0) -- (10mm,0); \draw
        [right] (10mm,0) arc (-180:-90:6mm); \draw (22mm,0) arc (0:-90:6mm);
        \draw [right,aux] (10mm,0) arc (180:90:6mm); \draw (22mm,0) arc
        (0:90:6mm); \draw [right, aux] (22mm,0) -- (27mm,0); \draw (27mm,0) --
        (32mm,0);
      \end{tikzpicture}
    \end{gathered}
    \label{eq:diff-sym-diagrams-3}
  \end{gather}
  \label{eq:diff-sym-diagrams}%
\end{subequations}
Relegating the explicit computations to the appendix, we find that
\begin{equation}
  G^{{\rm S},\text{1-loop}}_{nn}(p)
  = \frac{T^2\chi^2 \lambda^2\v[k]^2}{(\omega+iD\v[k]^2)^2}
  \frac{\tilde\sigma^2}{\sigma^2}\,\mathcal{I}_1(p)
  + \frac{T^2\chi^2 \lambda^2\v[k]^2}{(\omega-iD\v[k]^2)^2}
  \frac{\tilde\sigma^2}{\sigma^2}\,\mathcal{I}_1(-p).
  \label{eq:diff-S-1loop}
\end{equation}
It can be explicitly checked that, on imposing the KMS condition, these
corrections satisfy the fluctuation-dissipation theorem \eqref{fdt} along with
the retarded correlator corrections in \cref{eq:diff-R-1loop}. Finally, in
the $\omega=0$ limit, the symmetric correlator behaves as
\begin{equation}
  G^{{\rm S},\text{1-loop}}_{nn}(0,\v[k])
  = \frac{-2T^2\chi^2 \lambda^2}{D^2\v[k]^2}
  \frac{\tilde\sigma^2}{\sigma^2}\,\mathcal{I}_1(0,\v[k]),
  \label{eq:diff-S-1loop2}
\end{equation}
which survives the KMS limit $\tilde{\s} = \s$. This indicates that IR-singular non-analytic terms
are generated in the hydrostatic limit, as a manifestation of the lack of any
intrinsic mass scale in the tree-level propagators
\eqref{eq:diff-propagators}. This may appear surprising, given the earlier
discussion of thermal screening. However, the symmetric correlator is not
determined explicitly by the hydrostatic generating functional and does not have to meet the analyticity requirement. The leading IR-singularity, proportional to $1/{\v[k]}^2$, to the symmetric correlator is still given by the tree-level propagator. Also, note that the one-loop correction to the symmetric correlator in the static limit does not depend upon $\tilde{\lambda}$.


\section{Density correlation functions in diffusive hydrodynamics}
\label{sec:diff-hydro}

The discussion of charge diffusion above can be extended to the full
theory of relativistic hydrodynamics. We will again focus on the transport of a
single conserved charge, but consistently incorporate additional hydrodynamic
momentum modes in the thermal bath. We will develop an effective field theory
for \emph{diffusive hydrodynamics}, where the sound modes have been frozen and
the theory only contains diffusive charge and shear modes. This allows us to
compute the hydrodynamic correlators in a slightly simpler setting, while still
incorporating the non-trivial effects of momentum modes. Our goal is not to
provide a comprehensive analysis, but to test the spatial analyticity of
retarded charge density correlators in a system with additional gapless modes,
specifically those associated with momentum fluctuations.

\subsection{Stochastic diffusive hydrodynamics}
\label{sec:diff-hydro-action}

The classical dynamical equations for relativistic hydrodynamics are given by
charge and energy-momentum conservation,
\begin{equation}
  \dow_\mu J^\mu = 0, \qquad
  \dow_\mu T^{\mu\nu} = F^{\nu\lambda} J_\lambda,
\end{equation}
along with the constitutive relations expressing $J^\m$ and $T^{\m\n}$ in terms
of the hydrodynamic variables: temperature $T$, chemical potential $\m$, fluid
four-velocity $u^\mu$, and their derivatives. For instance, in the Landau frame
up to first order in derivatives we have\footnote{The symmetrization of indices follows the convention $\rm{A}^{(\m} \rm{B}^{\n)} = \half \left(\rm{A}^\m \rm{B}^\n + \rm{A}^\n \rm{B}^\m \right)$. Similarly, antisymmetrization follows the convention $\rm{A}^{[\m} \rm{B}^{\n]} = \half \left(\rm{A}^\m \rm{B}^\n - \rm{A}^\n \rm{B}^\m \right)$.}
\begin{align}
  J^\mu
  &= n(T,\mu)\, u^\mu
    - \sigma(T,\mu) \Delta^{\mu\lambda}
    \lb T \partial_\lambda \left(\frac{\mu}{T}\right)
    - F_{\lambda\nu} u^\nu \rb,\\
  T^{\mu\nu}
  &= \epsilon(T,\mu)\, u^\mu u^\nu + p(T,\mu)\,\Delta^{\mu\nu} \nn\\
  &\qquad
    - 2\eta(T,\mu) \lb \Delta^{\rho(\mu} \Delta^{\nu)\sigma} \dow_\rho u_\sigma
    - {\textstyle\frac{1}{d}} \Delta^{\mu\nu}\dow_\lambda u^\lambda
    \rb
    - \zeta(T,\mu) \Delta^{\mu\nu}
    \dow_\lambda u^\lambda.
    \label{eq:consti}
\end{align}
We have introduced the background gauge field $A_\mu$ coupled to $J^\mu$, and
the associated field strength $F_{\mu\nu} = 2\dow_{[\mu}A_{\nu]}$, but have
avoided introducing the metric source for $T^{\mu\nu}$ that we do not require
for our purposes.  Here $\Delta^{\mu\nu} = \eta^{\mu\nu} + u^\mu u^\nu$ is the
projector transverse to the fluid velocity. The thermodynamic pressure $p$,
energy density $\epsilon$, and charge density $n$ are functions of $T$ and
$\mu$, and related to each other via the thermodynamic relations
$dp = sdT + nd\mu$ and $\epsilon + p = Ts + \mu n$ for some entropy density
$s$. The charge conductivity $\sigma$, shear viscosity $\eta$, and bulk
viscosity $\zeta$ are non-negative functions of $T$ and $\mu$. We shall be
interested in fluctuations about an equilibrium state with $\mu = \mu_0$ at a
constant global temperature $T = T_0$ in the rest frame $u^\m = (1,\v[0])$.

The EFT for hydrodynamics is formulated in terms of a thermal vector $\beta^\mu$
and ``chemical shift'' field $\Lambda_\beta$, along with their stochastic noise
partners $X_a^\mu$ and $\varphi_a$. These are related to the aforementioned
hydrodynamic fields as $u^\mu/T = \beta^\mu$ and
$\mu/T = \Lambda_\beta + \beta^\mu A_{r\mu}$. Similar to the diffusive EFT, we
introduce the Schwinger-Keldysh double copies of background gauge fields
$A_{r\mu}$, $A_{a\mu}$. With all the ingredients in place, the full non-linear
effective action for relativistic hydrodynamics takes the
form~\cite{Kovtun:2014hpa}
\begin{align}
  \mathcal{L}
  &= J^\mu \lb \dow_\mu \varphi_a + A_{a\mu} + \lie_{X_a} A_{r\mu} \rb
    + i T \tilde\sigma \Delta^{\mu\nu}
    \big( \dow_\mu \varphi_a + A_{a\mu} + \lie_{X_a} A_{r\mu} \big)
    \big( \dow_\nu \varphi_a + A_{a\nu} + \lie_{X_a} A_{r\nu} \big) \nn\\
  &\qquad
    + T^{\mu\nu} \dow_\mu X_{a\nu}
    + i T \lb 2\tilde\eta \Delta^{\mu(\rho}\Delta^{\sigma)\nu}
    + (\tilde\zeta - {\textstyle\frac{2}{d}}\tilde\eta)
    \Delta^{\mu\nu}\Delta^{\rho\sigma} \rb
    \dow_\mu X_{a\nu} \dow_\rho X_{a\sigma}
    + \cdots,
    \label{Lhydro}
\end{align}
where $\lie_{X_a}$ denotes a Lie derivative along $X^\mu_a$. We have introduced
the stochastic coefficients $\tilde\sigma$, $\tilde\eta$, and $\tilde\zeta$,
which are also arbitrary non-negative functions of $T$ and $\mu$. We can recover
the previous diffusive model in \cref{L1} by simply setting $X_a^\mu = 0$ and
$u^\mu/T = \delta^\mu_t/T_0$. This is the most general effective action for
relativistic hydrodynamics, in Landau frame, compatible with the
Schwinger-Keldysh framework of~\cite{Crossley:2015evo}, truncated at
one-derivative order. We utilize the usual derivative counting scheme for
relativistic hydrodynamics: in the ``$r$'' sector we take
$u^\mu,T,\mu, A_{r\mu}\sim \mathcal{O}(\dow^0)$, while in the ``$a$'' sector we
have $X_a^\mu, \varphi_a \sim \mathcal{O}(\dow^0)$,
$A_{a\mu} \sim \mathcal{O}(\dow^1)$.

The dynamical KMS symmetry in \cref{eq:KMS-diff} generalizes to a covariant
version,
\begin{gather}
  \mu_0 \to -\mu_0, \qquad
  \beta^\mu(x) \to \beta^\mu(-x), \qquad
  \Lambda_\beta(x) \to -\Lambda_\beta(-x), \nn\\
  \varphi_a(x) \to \varphi_a(-x) + i\Lambda_\beta(-x), \qquad
  X^\mu_a(x) \to - X^\mu_a(-x)
  - i\lb \beta^\mu(-x) - T_0^{-1} \delta^\mu_t \rb, \nn\\
  A_{r\mu}(x) \to - A_{r\mu}(-x), \qquad
  A_{a\mu}(x) \to - A_{a\mu}(-x) - i(\lie_\beta A_{r\mu})(-x),
  \label{eq:KMS-hydro}
\end{gather}
where $\lie_\beta$ denotes a Lie-derivative along $\beta^\mu$. In addition to
the charge conjugation properties of various coefficients, the KMS symmetry
relates the stochastic fluctuation coefficients to the dissipative transport
coefficients via
\begin{equation}
 {\rm KMS:} \quad \tilde\sigma = \sigma, \qquad
  \tilde\eta = \eta, \qquad
  \tilde\zeta = \zeta.
  \label{kms2}
\end{equation}
However, we shall not implement these constraints for now to investigate the
role of KMS in ensuring the analyticity of hydrostatic retarded correlators.

Similar to our discussion in \cref{sec:diffusiveEFT}, we can expand the
Lagrangian order-by-order in the hydrodynamic fluctuations
$\delta\beta^\mu = \beta^\mu - \delta^\mu_t/T_0$,
$\delta\Lambda_\beta = \Lambda_\beta - \mu_0/T_0$, and associated noise
$X_a^\mu$, $\varphi_a$. The full theory turns out to be quite complicated in
practice. To make our lives simpler, we will focus on a sub-sector, which we
call \textit{diffusive hydrodynamics}, where the temperature and the
longitudinal velocity fluctuations have been frozen and so can be ignored. This
is similar to the well-known ``incompressible limit'' of non-relativistic
hydrodynamics and amounts to ignoring the sound modes and only focusing on the
diffusive modes. We take
\begin{equation}
  \delta\beta^t = 0, \qquad
  \dow_i \delta\beta^i = 0, \qquad
  X^t_{a} = 0, \qquad
  \dow_iX_{a}^i =0.
  \label{eq:incompressibility}
\end{equation}
The last two conditions are required for consistency with the KMS condition
\eqref{eq:KMS-hydro}. This removes the energy conservation equation and the
trace of the spatial stress tensor from the effective action. In terms of the
original hydrodynamic variables, the first two conditions imply\footnote{If we
  were to decompose the relativistic fluid velocity according to
  $u^\mu = (1,\v[v])/\sqrt{1-\v[v]^2}$, these conditions will imply
  $\dow_i v^i = 0$ exactly, making contact with the incompressible limit of
  hydrodynamics.}
\begin{equation}
  \delta T \equiv T - T_0 = \frac{T_0}{2} \v[u]^2 + \ldots, \qquad
  \dow_i u^i = 0 + \ldots,
\end{equation}
where $\v[u]^2 = u^i u_i$. We have suppressed terms that are cubic or higher
order in $\delta\beta^i$. To impose these conditions in the EFT, we include
additional constraint terms in the effective action
\begin{equation}
  \mathcal{L}_{\text{fixing}}
  = \Psi_a \dow_i \delta\beta^i
  + \Psi_r \dow_i X^i_a,
\end{equation}
for arbitrary Lagrange multipliers $\Psi_{r}$, $\Psi_a$. For consistency with
the KMS condition \eqref{eq:KMS-hydro}, these multipliers must transform as
\begin{equation}
  \Psi_r(x) \to \Psi_r(-x), \qquad
  \Psi_a(x) \to -\Psi_a(-x) + i\Psi_r(-x).
\end{equation}

\subsection{Linear expansion and interactions}
\label{sec:hydroaction_expansion}

It turns out to be more convenient to work with density fluctuations
$\delta n = J^t|_{A_r=0} - n(\mu_0) + 2i T \tilde\sigma u^i \dow_i \varphi_a$
and momentum fluctuations
$\pi^i = T^{ti} + 2 i T\tilde{\eta} (u^k \dow_k X^i_{a} + u_k \dow^i
X^k_a)$.\footnote{These definitions of $\delta n$ and $\pi^i$, including the
  imaginary parts, are the true charge and momentum densities that are conserved
  in the presence of stochastic fluctuations, unlike $J^0$ and $T^{0i}$ that are
  only conserved classically; see \cref{details} for details.}  The free part of
the effective Lagrangian \eqref{Lhydro}, along with the Lagrange multiplier
terms, takes the form
\begin{align}
  \mathcal{L}_{\text{free}}
  &= - \varphi_a \lb \dow_t - D \N^2 \rb \delta n
    + i T \tilde\sigma\, \dow_i \varphi_a \dow^i \varphi_a
    - X^i_{a} \lb \dow_t - \gamma_\eta \N^2 \rb \pi_i 
    + i T \tilde \eta\, \dow_k  X^j_{a} \dow^k X_{aj} \nn\\
  &\qquad
    + \frac{1}{Tw} \Psi_a \dow_i \pi^i
    + \Psi_r \dow_i X_a^i.
\label{freelang}
\end{align}
The explicit calculational details can be found in appendix \ref{details}. As before,
$D = \sigma/\chi$ is the charge diffusion constant with
$\chi = \dow n/\dow\mu|_T$ being the charge susceptibility, while
$\gamma_\eta = \eta/w$ is the shear diffusion constant with $w = \epsilon+p$
being the enthalpy density. We still have the same tree-level propagators in the
charge sector, \cref{eq:diff-propagators}, but we also find new momentum
dependent propagators given by
\begin{align}
  \langle \pi^i(p)X^j_a(-p)\rangle_0
  &= \lb \delta^{ij} - \frac{k^i k^j}{\v[k]^2} \rb
    \frac{1}{\omega + i\gamma_\eta\v[k]^2},
  && \raisebox{-3mm}{\begin{tikzpicture}[thick]
    \draw [mom] (0mm,0)  -- (10mm,0);
    \draw [right] (0mm,0)  -- (10mm,0);
    \draw [coil] (10mm,0)  -- (20mm,0);
    \node at (10mm,-3mm) {$p$};
	\node at (0mm,-2.5mm) {$i$};
	\node at (20mm,-2.5mm) {$j$};
  \end{tikzpicture}} \nn\\
  \langle X^i_a(p)\pi^j(-p)\rangle_0
  &= \lb \delta^{ij} - \frac{k^i k^j}{\v[k]^2} \rb
    \frac{-1}{\omega - i\gamma_\eta\v[k]^2},
  && \raisebox{-3mm}{\begin{tikzpicture}[thick]
      \draw [coil,right] (0mm,0)  -- (10mm,0);
      \draw [mom] (10mm,0)  -- (20mm,0);
      \node at (10mm,-3mm) {$p$};
	\node at (0mm,-2.5mm) {$i$};
	\node at (20mm,-2.5mm) {$j$};
    \end{tikzpicture}} \nn\\
  \langle \pi^i(p)\pi^j(-p)\rangle_0
  &= \lb \delta^{ij} - \frac{k^i k^j}{\v[k]^2} \rb
    \frac{2T\tilde\eta\v[k]^2}{\omega^2 + \gamma_\eta^2\v[k]^4},
  && \raisebox{-3mm}{\begin{tikzpicture}[thick]
      \draw [right] (0mm,0)  -- (10mm,0);
      \draw [mom] (0mm,0)  -- (10mm,0);
      \draw [mom] (10mm,0)  -- (20mm,0);
      \node at (10mm,-3mm) {$p$};
	\node at (0mm,-2.5mm) {$i$};
	\node at (20mm,-2.5mm) {$j$};
    \end{tikzpicture}} \nn\\
  \langle X_a^i(p)X_a^j(-p)\rangle_0
  &= 0.
     \label{eq:mom-propagators}
\end{align}
We have denoted $\pi^i$ with a bold and $X^i_a$ with a coiled line. All the
cross-propagators between the momentum and the charge sectors are zero. This is
a feature of the simplified diffusive hydrodynamic theory eq.\ \eqref{eq:incompressibility}, and will not be the case in full hydrodynamics in the presence of sound modes. These propagators are
made transverse to momentum by the presence of the Lagrange multipliers. The
propagators for the multipliers themselves can be computed, but they do not have
any physical poles and drop out for our case of interest.

At the next order in fluctuations, we have the three-point interaction
Lagrangian
\begin{align}
  \mathcal{L}^{\text{3pt}}_{\text{int}}
  &= \half \lambda \delta n^2 \nabla^2 \varphi_a
    + iT\chi\tilde\lambda\, \delta n\, \dow_i\varphi_a \dow^i\varphi_a
    + \frac{1}{w} \pi^i \delta n\, \dow_i\varphi_a
    + \frac{1}{w} \pi^i \pi^j \dow_i X_{aj}  \nn\\
  &\qquad
    + \half \lambda_\pi \v[\pi]^2 \nabla^2 \varphi_a
    - \psi\, \pi^j \pi^i \dow_i\dow_j\varphi_a 
    - i Tw \tilde\psi
    \lb \pi_j \dow^iX_{a}^j + \pi_j \dow^jX_{a}^i \rb
    \dow_i\varphi_a \nn\\
  &\qquad
    - \gamma\,\delta n\, \pi^{j} \nabla^2X_{aj}
    - 2\theta\,\delta n\, \dow^{(i}\pi^{j)} \dow_iX_{aj}
    + 2i Tw\tilde\theta\, \delta n\, \dow^{(i}X^{j)}_{a} \dow_iX_{aj} \nn\\
  &\qquad
    - \frac{\chi_\epsilon}{T\chi w^2} \Psi_a \pi^i \dow_i\delta n,
    \label{Lhydroint}
\end{align}
which generalizes \cref{L1int}. Various coupling constants are defined as
\begin{gather}
  \chi_\epsilon = \frac{\dow w}{\dow\mu}, \qquad
  \lambda
  = \frac{1}{\chi}\frac{\dow D}{\dow \mu}, \qquad
  \tilde\lambda
  = \frac{1}{\chi^2} \frac{\dow\tilde\sigma}{\dow\mu}, \qquad
  \lambda_\pi
  = - \frac{1}{w^2} \lb \chi_\epsilon D + n \gamma_\eta \rb, \nn\\
  \psi = \frac{n\eta}{w^3}, \quad
  \tilde\psi = \frac{2n\tilde\eta}{w^3}, \qquad
  \gamma = \frac{\eta(n + \chi_\epsilon)}{\chi w^2}, \qquad
  \theta = \frac{1}{\chi w} \lb \frac{\eta n}{w}
  + \frac{\dow\eta}{\dow\mu} \rb, \quad
  \tilde\theta = \frac{1}{\chi w} \frac{\dow\tilde\eta}{\dow\mu}.
\end{gather}
The vertices arising from the first two interactions in \cref{Lhydroint} are
the same as given in \cref{eq:vertices-diff}. In addition, we get six more
vertices
\begin{align}
  &\raisebox{-9mm}{\begin{tikzpicture}[thick]
      \draw [aux,left] (-10mm,0)  -- (0,0);
      \node at (-7mm,-3mm) {$p$};
      \draw [right] (0,0)  -- (7mm,7mm);
      \node at (3mm,7mm) {$p'$};
      \draw [mom] (0,0)  -- (7mm,-7mm);
      \draw [right] (0,0)  -- (7mm,-7mm);
      \node at (3mm,-7mm) {$p''$};
      \node at (7mm,-3mm) {$i$};
    \end{tikzpicture}}
    \qquad \frac{-1}{w} k_i, \qquad\qquad
  &&\raisebox{-9mm}{\begin{tikzpicture}[thick]
      \draw [left] (-10mm,0)  -- (-9.9mm,0);
      \draw [coil] (-10mm,0)  -- (0,0);
      \node at (-7mm,-3mm) {$p$};
      \node at (-7mm,3mm) {$i$};
      \draw [mom] (0,0)  -- (7mm,7mm);
      \draw [right] (0,0)  -- (7mm,7mm);
      \node at (3mm,7mm) {$p'$};
      \node at (7mm,3mm) {$j$};
      \draw [mom] (0,0)  -- (7mm,-7mm);
      \draw [right] (0,0)  -- (7mm,-7mm);
      \node at (3mm,-7mm) {$p''$};
      \node at (7mm,-3mm) {$k$};
    \end{tikzpicture}}
    \qquad \frac{-1}{2w} (k_j \delta_{ik} + k_k \delta_{ij}), \qquad\qquad \nn\\
  &\raisebox{-9mm}{\begin{tikzpicture}[thick]
      \draw [aux,left] (-10mm,0)  -- (0,0);
      \node at (-7mm,-3mm) {$p$};
      \draw [mom] (0,0)  -- (7mm,7mm);
      \draw [right] (0,0)  -- (7mm,7mm);
      \node at (3mm,7mm) {$p'$};
      \node at (7mm,3mm) {$i$};
      \draw [mom] (0,0)  -- (7mm,-7mm);
      \draw [right] (0,0)  -- (7mm,-7mm);
      \node at (3mm,-7mm) {$p''$};
      \node at (7mm,-3mm) {$j$};
    \end{tikzpicture}}
  \qquad \frac{-i}{2}\lambda_\pi \delta_{ij} \v[k]^2
    + i\psi\, k_i k_j, \qquad
  &&\raisebox{-9mm}{\begin{tikzpicture}[thick]
      \draw [left] (-10mm,0)  -- (-9.9mm,0);
      \draw [coil] (-10mm,0)  -- (0,0);
      \node at (-7mm,-3mm) {$p$};
      \node at (-7mm,3mm) {$i$};
      \draw [right] (0,0)  -- (7mm,7mm);
      \node at (3mm,7mm) {$p'$};
      \draw [mom] (0,0)  -- (7mm,-7mm);
      \draw [right] (0,0)  -- (7mm,-7mm);
      \node at (3mm,-7mm) {$p''$};
      \node at (7mm,-3mm) {$j$};
    \end{tikzpicture}}
     \qquad i\gamma \delta_{ij} \v[k]^2
     + i\theta (\delta_{ij} \v[k]\cdot\v[k]'' + k_j k''_i), \nn\\
  &\raisebox{-9mm}{\begin{tikzpicture}[thick]
      \draw [mom] (-10mm,0)  -- (0,0);
      \draw [left] (-10mm,0)  -- (0,0);
      \node at (-7mm,-3mm) {$p$};
      \node at (-7mm,3mm) {$i$};
      \draw [coil,right] (0,0)  -- (7mm,7mm);
      \node at (3mm,7mm) {$p'$};
      \node at (7mm,3mm) {$j$};
      \draw [aux,right] (0,0)  -- (7mm,-7mm);
      \node at (3mm,-7mm) {$p''$};
    \end{tikzpicture}}
    \qquad -Tw \tilde\psi (\delta_{ij} \v[k]'\cdot \v[k]'' + k'_i k''_j ),
  &&\raisebox{-9mm}{\begin{tikzpicture}[thick]
      \draw [left] (-10mm,0)  -- (0,0);
      \node at (-7mm,-3mm) {$p$};
      \draw [coil,right] (0,0)  -- (7mm,7mm);
      \node at (3mm,7mm) {$p'$};
      \node at (7mm,3mm) {$i$};
      \draw [coil,right] (0,0)  -- (7mm,-7mm);
      \node at (3mm,-7mm) {$p''$};
      \node at (7mm,-3mm) {$j$};
    \end{tikzpicture}}
    \qquad Tw\tilde\theta\, (\delta_{ij}\v[k]'\cdot \v[k]'' + k'_j k''_i).
\end{align}
There is another interaction vertex involving the Lagrange multiplier $\Psi_a$.
However, all the loop diagrams involving this vertex include a purely
retarded/advanced loop integral and hence do not contribute to the problem at
hand.

Finally, we need the coupling structure to the background fields. Turning on
only the $A_{rt}$, $A_{at}$ components of the gauge fields, we find
\begin{align}
  \mathcal{L}_{\text{source}}^{\text{2pt}}
  &= \delta n\, A_{at}
    + \chi \dow_t\varphi_a A_{rt}
    + \chi A_{rt}A_{at}, \nn\\
  \mathcal{L}_{\text{source}}^{\text{3pt}}
  &= \frac{\chi'}{\chi} \delta n\, A_{rt}A_{at}
    + \half\chi' \dow_t\varphi_a A_{rt}^2
    + \half\chi' A_{rt}^2A_{at} \nn\\
  &\qquad
    + \lb \frac{\chi'}{\chi} \delta n\, \dow_t\varphi_a
    - \frac{1}{\chi} \frac{\dow\sigma}{\dow\mu}
    \partial^i \delta n \dow_i\varphi_a
    + iT \frac{\dow\tilde\sigma}{\dow\mu} \dow_i \varphi_a \dow^i \varphi_a
    \rb A_{rt}
    + \lb \frac{\chi}{w} \pi^i \dow_i\varphi_a
    - X^i_{a} \dow_i\delta n
    \rb A_{rt}  \nn\\
  &\qquad
    + \lb \frac{\chi_\epsilon}{w}  \pi_i \dow_t X_{a}^i
    - \frac{2}{w}\frac{\dow\eta}{\dow \mu}  \dow^{(i} \pi^{j)}  \dow_i X_{aj}
    + 2i T \frac{\dow\tilde\eta}{\dow\mu}
    \dow^{(i} X^{j)}_{a} \dow_i X_{aj} \rb A_{rt}.
\label{sourlang}
\end{align}
See the appendix \ref{details} for details of the derivation. Most of these are
the same as the ones found in the diffusive case in \cref{L1source}, but we do
find four additional background interaction vertices, given by
\begin{align}
  &\raisebox{-9mm}{\begin{tikzpicture}[thick]
      \draw [right,dotted] (0,0)  -- (9mm,0);
      \node at (7mm,-3mm) {$p$};
      \draw [aux,left] (-7mm,7mm) -- (0,0);
      \node at (-3mm,8mm) {$p'$};
      \draw [left] (-7mm,-7mm) -- (0,0);
      \draw [mom] (-7mm,-7mm) -- (0,0);
      \node at (-3mm,-7mm) {$p''$};
      \node at (-7mm,-3mm) {$i$};
    \end{tikzpicture}}\qquad
    \frac{-\chi}{w} k'_i, 
  &&\raisebox{-9mm}{\begin{tikzpicture}[thick]
      \draw [right,dotted] (0,0)  -- (9mm,0);
      \node at (7mm,-3mm) {$p$};
      \draw [left] (-7mm,7mm) -- (-6.9mm,6.9mm);
      \draw [coil] (-7mm,7mm) -- (0,0);
      \node at (-3mm,8mm) {$p'$};
      \node at (-7mm,3mm) {$i$};
      \draw [left] (-7mm,-7mm) -- (0,0);
      \node at (-3mm,-7mm) {$p''$};
    \end{tikzpicture}}\qquad
     k''_i, \nn\\
  &\raisebox{-9mm}{\begin{tikzpicture}[thick]
      \draw [right,dotted] (0,0)  -- (9mm,0);
      \node at (7mm,-3mm) {$p$};
      \draw [left] (-7mm,7mm) -- (-6.9mm,6.9mm);
      \draw [coil] (-7mm,7mm) -- (0,0);
      \node at (-3mm,8mm) {$p'$};
      \node at (-7mm,3mm) {$i$};
      \draw [left] (-7mm,-7mm) -- (0,0);
      \draw [mom] (-7mm,-7mm) -- (0,0);
      \node at (-3mm,-7mm) {$p''$};
      \node at (-7mm,-3mm) {$j$};
    \end{tikzpicture}}\quad
    \frac{\chi_\epsilon}{w} \omega'\delta_{ij}
    + \frac{i}{w} \frac{\dow\eta}{\dow\mu}
    (\delta_{ij}\v[k]'\cdot\v[k]'' + k'_jk''_i), 
  &&\raisebox{-9mm}{\begin{tikzpicture}[thick]
      \draw [right,dotted] (0,0)  -- (9mm,0);
      \node at (7mm,-3mm) {$p$};
      \node at (-7mm,3mm) {$i$};
      \draw [left] (-7mm,7mm) -- (-6.9mm,6.9mm);
      \draw [coil] (-7mm,7mm) -- (0,0);
      \node at (-3mm,8mm) {$p'$};
      \node at (-7mm,3mm) {$i$};
      \draw [coil] (-7mm,-7mm) -- (0,0);
      \draw [left] (-7mm,-7mm) -- (-6.9mm,-6.9mm);
      \node at (-3mm,-7mm) {$p''$};
      \node at (-7mm,-3mm) {$j$};
    \end{tikzpicture}}\qquad
     T \frac{\dow\tilde\eta}{\dow\mu}
     (\delta_{ij}\v[k]'\cdot\v[k]'' + k'_jk''_i),
\end{align}

\subsection{One-loop corrections from momentum fluctuations}
We can use this theory to work out the corrections to the density correlation
functions arising from momentum fluctuations. The variational formulas in
\cref{eq:variational_formulas} are still valid, but with the modified generating
functional (see \cref{foot:jacobian})
\begin{equation}
  \exp W[A_{r\mu},A_{a\mu}]
  = \int \mathcal{D}\delta n\mathcal{D}\varphi_a
  \mathcal{D}\pi^i\mathcal{D}X_a^i
  \mathcal{D}\Psi_r\mathcal{D}\Psi_a
  \exp\lb i \int \df^{d+1} x\,\mathcal{L} \rb.
\end{equation} 
This still leads to the same tree-level two-point correlators as reported in
\cref{eq:diff-propagators}.

\subsubsection{Retarded correlation function}

For the retarded function at one-loop order in diffusive hydrodynamics, we still get contributions from the previous diagrams in \cref{eq:diff-ret-diagrams}. In addition, we have the following three one-loop diagrams
involving momentum fluctuations contributing to the $\langle \delta n\varphi_a\rangle$ propagator:
\begin{subequations}
  \begin{equation}
    \begin{aligned}
      & \begin{tikzpicture}[thick]
        \draw [right] (0mm,0)  -- (5mm,0);
        \draw [aux] (5mm,0)  -- (10mm,0);
        \draw [right] (10mm,0) arc (-180:-90:6mm);
        \draw [mom] (10mm,0) arc (-180:0:6mm);
        \draw [right] (10mm,0) arc (180:90:6mm);
        \draw [aux] (22mm,0) arc (0:90:6mm);
        \draw [right] (22mm,0)  -- (27mm,0);
        \draw [aux] (27mm,0)  -- (32mm,0);
      \end{tikzpicture} \qquad
      && \begin{tikzpicture}[thick]
        \draw [right] (0mm,0)  -- (5mm,0);
        \draw [aux] (5mm,0)  -- (10mm,0);
        \draw [right] (10mm,0) arc (-180:-90:6mm);
        \draw [mom] (10mm,0) arc (-180:0:6mm);
        \draw [right] (10mm,0) arc (180:90:6mm);
        \draw [mom] (10mm,0) arc (180:90:6mm);
        \draw [coil] (22mm,0) arc (0:90:6mm);
        \draw [right] (22mm,0)  -- (27mm,0);
        \draw [aux] (27mm,0)  -- (32mm,0);
      \end{tikzpicture} \qquad
      && \begin{tikzpicture}[thick]
        \draw [right] (0mm,0)  -- (5mm,0);
        \draw [aux] (5mm,0)  -- (10mm,0);
        \draw [right] (10mm,0) arc (-180:-90:6mm);
        \draw [mom] (10mm,0) arc (-180:-90:6mm);
        \draw [coil] (22mm,0) arc (0:-90:6mm);
        \draw [right] (10mm,0) arc (180:90:6mm);
        \draw [mom] (10mm,0) arc (180:90:6mm);
        \draw [coil] (22mm,0) arc (0:90:6mm);
        \draw [right] (22mm,0)  -- (27mm,0);
        \draw [aux] (27mm,0)  -- (32mm,0);
      \end{tikzpicture}
    \end{aligned}
    \label{eq:hydro-ret-diagrams-prop}
  \end{equation}
With the incorporation of momentum fluctuations, we also have four new background coupling diagrams:
  \begin{equation}
    \begin{aligned}
      & \begin{tikzpicture}[thick]
        \draw [right] (0mm,0)  -- (5mm,0);
        \draw [aux] (5mm,0)  -- (10mm,0);
        \draw [right] (10mm,0) arc (-180:-90:6mm);
        \draw (22mm,0) arc (0:-90:6mm);
        \draw [mom] (10mm,0) arc (180:90:6mm);
        \draw [right] (10mm,0) arc (180:90:6mm);
        \draw [coil] (22mm,0) arc (0:90:6mm);
        \draw [right, dotted] (22mm,0)  -- (28mm,0);
      \end{tikzpicture}\quad
      && \begin{tikzpicture}[thick]
        \draw [right] (0mm,0)  -- (5mm,0);
        \draw [aux] (5mm,0)  -- (10mm,0);
        \draw [right] (10mm,0) arc (-180:-90:6mm);
        \draw [mom] (10mm,0) arc (-180:0:6mm);
        \draw [right] (10mm,0) arc (180:90:6mm);
        \draw [aux] (22mm,0) arc (0:90:6mm);
        \draw [right, dotted] (22mm,0)  -- (28mm,0);
      \end{tikzpicture}\quad
      && \begin{tikzpicture}[thick]
        \draw [right] (0mm,0)  -- (5mm,0);
        \draw [aux] (5mm,0)  -- (10mm,0);
        \draw [right] (10mm,0) arc (-180:-90:6mm);
        \draw [mom] (10mm,0) arc (-180:0:6mm);
        \draw [right] (10mm,0) arc (180:90:6mm);
        \draw [mom] (10mm,0) arc (180:90:6mm);
        \draw [coil] (22mm,0) arc (0:90:6mm);
        \draw [right, dotted] (22mm,0)  -- (28mm,0);
      \end{tikzpicture}\quad
      && \begin{tikzpicture}[thick]
        \draw [right] (0mm,0)  -- (5mm,0);
        \draw [aux] (5mm,0)  -- (10mm,0);
        \draw [right] (10mm,0) arc (-180:-90:6mm);
        \draw [mom] (10mm,0) arc (-180:-90:6mm);
        \draw [coil] (22mm,0) arc (0:-90:6mm);
        \draw [right] (10mm,0) arc (180:90:6mm);
        \draw [mom] (10mm,0) arc (180:90:6mm);
        \draw [coil] (22mm,0) arc (0:90:6mm);
        \draw [right, dotted] (22mm,0)  -- (28mm,0);
      \end{tikzpicture}
    \end{aligned}
    \label{eq:hydro-ret-diagrams-back}
  \end{equation}
  \label{eq:hydro-ret-diagrams}%
\end{subequations}
We have ignored the possible diagrams which involve a purely retarded loop and
hence do not contribute to the final result. The first diagram in
\cref{eq:hydro-ret-diagrams-prop} and the first two diagrams in
\cref{eq:hydro-ret-diagrams-back} involve leading derivative non-dissipative
couplings and dominate in the IR. The remaining four diagrams here and the ones
in \cref{eq:diff-ret-diagrams} lead to subleading corrections in $\v[k]^2$. The
explicit computation of these diagrams can be found in the appendix
\ref{app:loop_details}. Combining the result from
\cref{eq:diff-ret-diagrams,eq:hydro-ret-diagrams}, we find that the full
one-loop correction to the retarded density correlation function in diffusive
hydrodynamics is given by
\begin{align}
  G^{\text{R},\text{1-loop}}_{nn}(p)
  &= \frac{-T\v[k]^2}{(\omega +iD\v[k]^2)^2}
    \Bigg[
    \frac{\chi}{w}\lb i \omega\frac{\tilde\sigma}{\sigma}
    + D\v[k]^2 \lb \frac{\tilde\eta}{\eta}
    - \frac{\tilde\sigma}{\sigma}\rb  \rb \mathcal{I}_0(p) \nn\\
  &\qquad
    - \lambda
    \lb  \frac{\tilde\sigma}{\sigma} i\omega \chi^2  \lambda
    - (i\omega - D\v[k]^2) \sigma \frac{\dow(\tilde\sigma/\sigma)}{\dow\mu}
     \rb \mathcal{I}_1(p) \nn\\
  &\qquad
    + i\omega \frac{n_0\tilde\eta}{w} \lb
    \lambda_\pi\mathcal{J}_1(p) + \lambda_\pi\mathcal{J}_2(p)
    - \psi\mathcal{J}_3(p) -\psi\mathcal{J}_4(p)
    \rb
    \nn\\
  &\qquad
    + D\v[k]^2 \lb
    \frac{\tilde\eta}{w} \chi_\epsilon
    \lb \lambda_\pi \mathcal{J}_1(p) - \psi \mathcal{J}_3(p)
    \rb
    + \eta \frac{\dow(\tilde\eta/\eta)}{\dow\mu}
    \lb \lambda_\pi \mathcal{J}_2(p) - \psi \mathcal{J}_4(p) \rb \rb \nn\\
  &\qquad
    - \half \frac{\tilde\eta}{w}\chi_\epsilon(\lambda_\pi - \psi)
       (i\omega - D\v[k]^2) \mathcal{I}_3(0,\v[k]) \Bigg].
    \label{eq:GR-Hydro}
\end{align}
The integral $\mathcal{I}_1(p)$ is defined in \cref{eq:I1}. We define three more
similar integrals via
\begin{align}
  \mathcal{I}_0(p)
  &= \frac{1}{\v[k]^2} \int^\Lambda \frac{\df^{d}\v[k]'}{(2\pi)^d}
    \frac{\v[k]^2 - (\v[k]\cdot\v[k]')^2/\v[k]'^2}{
    i\omega - \gamma_\eta \v[k]'^2 - D(\v[k]{-}\v[k]')^2} \nn\\
  &\qquad\overset{d=3}{=}
    \frac{|\v[k]|}{32 \pi D}
    \lb 
    \frac{2\sqrt{\gamma_\eta D}(D-\gamma_\eta)}{(\gamma_\eta +D)^2} 
    + \arccos\lb\frac{\gamma_\eta-D}{\gamma_\eta +D}\rb \rb
    + \mathcal{O}(\omega), \nn\\
  \mathcal{I}_2(p)
  &= \frac{\v[k]^2}{2} \int^\Lambda \frac{\df^{d}\v[k]'}{(2\pi)^d}
  \frac{1}{i\omega - \gamma_\eta (\v[k]'^2 + (\v[k]-\v[k]')^2)}
  \overset{d=3}{=}
  \frac{|\v[k]|^3}{32 \pi D}  + \mathcal{O}(\omega), \nn\\
  \mathcal{I}_3(p)
  &= \frac{\v[k]^2}{2} \int^\Lambda \frac{\df^{d}\v[k]'}{(2\pi)^d}
  \frac{-\v[k]^2/\v[k]'^2}{i\omega - \gamma_\eta (\v[k]'^2 + (\v[k]-\v[k]')^2)}
  \overset{d=3}{=}
    \frac{|\v[k]|^3}{32\gamma_\eta}
    + \frac{i \omega |\v[k]|}{16 \pi \gamma_\eta^2}
    +\mathcal{O}(\omega^2).
    \label{eq:I0}
\end{align}
For reference, we have computed these integrals explicitly in $d=3$ for small
$\omega$, with a hard momentum cutoff $\L$. However, the analyticity results do not
depend on the explicit form of the integrals and are valid in arbitrary $d$. In
terms of these, the $\mathcal{J}_{1,2,3,4}(p)$ integrals appearing in
\cref{eq:GR-Hydro} are defined as
\begin{align}
  \mathcal{J}_1(p)
  &= (d-2) \frac{i\omega}{\gamma_\eta\v[k]^2} \mathcal{I}_2(p)
    - \frac{(i\omega - \gamma_\eta\v[k]^2)^2}{2\gamma_\eta^2\v[k]^4} \mathcal{I}_3(p), \nn\\
    \mathcal{J}_2(p)
  &= -(d-3) \frac{i\omega - \gamma_\eta\v[k]^2}{\gamma_\eta\v[k]^2}
    \mathcal{I}_2(p)
  + \frac{\gamma_\eta \v[k]^2}{i\omega}
    \lb \frac{(i\omega - \gamma_\eta\v[k]^2)^3}{\gamma_\eta^3 \v[k]^6}
    \mathcal{I}_3(p)
    + \mathcal{I}_3(0,\v[k]) \rb, \nn\\
  \mathcal{J}_3(p)
  &= \frac{i\omega (i\omega - \gamma_\eta\v[k]^2)}{\gamma_\eta^2\v[k]^4} \mathcal{I}_2(p)
    + \frac{(i\omega - \gamma_\eta\v[k]^2)^3}{2\gamma_\eta^3\v[k]^6} \mathcal{I}_3(p), \nn\\
  \mathcal{J}_4(p)
  &= \lb
    \frac{2\omega^2}{\gamma_\eta^2\v[k]^4}
    - \frac{3d-1}{d\v[k]^2} \lb \v[k]^2
    - \frac{2i\omega}{\gamma_\eta} \rb
    \rb
    \mathcal{I}_2(p)
    - \frac{\gamma_\eta \v[k]^2}{i\omega}
    \lb \frac{(i\omega - \gamma_\eta\v[k]^2)^4}{\gamma_\eta^4 \v[k]^8}
    \mathcal{I}_3(p)
    - \mathcal{I}_3(0,\v[k])  \rb.
    \label{eq:Jexpansion}
\end{align}
While $\mathcal{J}_2(p)$ and $\mathcal{J}_4(p)$ have a $1/\omega$ appearing in
their definitions, they are perfectly regular in the $\omega\to 0$ limit.

These expressions are fairly involved. However, if we set $\omega=0$, the
one-loop correction to the retarded correlation function reduces to
\begin{align}
  G^{\text{R},\text{1-loop}}_{nn}(0,\bf{k})
  &= \frac{T}{D}
    \Bigg[\frac{\chi}{w}
    \lb \frac{\tilde\eta}{\eta}
    - \frac{\tilde\sigma}{\sigma}\rb \mathcal{I}_0(0,\v[k])  \nn\\
  &\qquad\qquad
    - \lambda \sigma \frac{\dow(\tilde\sigma/\sigma)}{\dow\mu}
    \mathcal{I}_1(0,\v[k])
    + \eta \frac{\dow(\tilde\eta/\eta)}{\dow\mu}
    \lb \lambda_\pi \mathcal{J}_2(0,\v[k])
    - \psi \mathcal{J}_4(0,\v[k]) \rb \Bigg], \nn\\
  &\overset{\text{KMS}}{=} 0.
    \label{eq:GR-Hydro-0}
\end{align}
Hence, we see that the non-analyticities drop out of the retarded correlator in the hydrostatic limit, even after the incorporation of momentum fluctuations, provided the KMS condition is imposed.

\subsubsection{Symmetric correlation function}

We can repeat the same procedure for the symmetric correlation function as
well. In addition to the 7 diagrams in \cref{eq:diff-sym-diagrams}, we find 10
more in the presence of momentum fluctuations:
\begin{subequations}
  \begin{gather}
    \begin{gathered}
      \begin{tikzpicture}[thick]
        \draw [right] (0mm,0)  -- (5mm,0);
        \draw [aux] (5mm,0)  -- (10mm,0);
        \draw [right] (10mm,0) arc (-180:-90:6mm);
        \draw [mom] (10mm,0) arc (-180:0:6mm);
        \draw [right] (10mm,0) arc (180:90:6mm);
        \draw [aux] (22mm,0) arc (0:90:6mm);
        \draw [right] (22mm,0)  -- (27mm,0);
        \draw (27mm,0)  -- (32mm,0);
      \end{tikzpicture}\qquad
      \begin{tikzpicture}[thick]
        \draw [right] (0mm,0)  -- (5mm,0);
        \draw [aux] (5mm,0)  -- (10mm,0);
        \draw [right] (10mm,0) arc (-180:-90:6mm);
        \draw [mom] (10mm,0) arc (-180:0:6mm);
        \draw [right] (10mm,0) arc (180:90:6mm);
        \draw [mom] (10mm,0) arc (180:90:6mm);
        \draw [coil] (22mm,0) arc (0:90:6mm);
        \draw [right] (22mm,0)  -- (27mm,0);
        \draw (27mm,0)  -- (32mm,0);
      \end{tikzpicture}\qquad
      \begin{tikzpicture}[thick]
        \draw [right] (0mm,0)  -- (5mm,0);
        \draw [aux] (5mm,0)  -- (10mm,0);
        \draw [right] (10mm,0) arc (-180:-90:6mm);
        \draw [mom] (10mm,0) arc (-180:-90:6mm);
        \draw [coil] (22mm,0) arc (0:-90:6mm);
        \draw [right] (10mm,0) arc (180:90:6mm);
        \draw [mom] (10mm,0) arc (180:90:6mm);
        \draw [coil] (22mm,0) arc (0:90:6mm);
        \draw [right] (22mm,0)  -- (27mm,0);
        \draw (27mm,0)  -- (32mm,0);
      \end{tikzpicture}
    \end{gathered}
    \label{eq:hydro-sym-diagrams-1}\\
    \begin{gathered}
      \begin{tikzpicture}[thick]
        \draw [right] (0mm,0)  -- (5mm,0);
        \draw (5mm,0)  -- (10mm,0);
        \draw [right] (10mm,0) arc (-180:-90:6mm);
        \draw [mom] (10mm,0) arc (-180:0:6mm);
        \draw [aux,right] (10mm,0) arc (180:90:6mm);
        \draw (22mm,0) arc (0:90:6mm);
        \draw [aux,right] (22mm,0)  -- (27mm,0);
        \draw (27mm,0)  -- (32mm,0);
      \end{tikzpicture}\qquad
      \begin{tikzpicture}[thick]
        \draw [right] (0mm,0)  -- (5mm,0);
        \draw (5mm,0)  -- (10mm,0);
        \draw [right] (10mm,0) arc (-180:-90:6mm);
        \draw [mom] (10mm,0) arc (-180:0:6mm);
        \draw [right] (16mm,6mm) -- (16.1mm,6mm);
        \draw [coil] (10mm,0) arc (180:90:6mm);
        \draw [mom] (22mm,0) arc (0:90:6mm);
        \draw [aux,right] (22mm,0)  -- (27mm,0);
        \draw (27mm,0)  -- (32mm,0);
      \end{tikzpicture}\qquad
      \begin{tikzpicture}[thick]
        \draw [right] (0mm,0)  -- (5mm,0);
        \draw (5mm,0)  -- (10mm,0);
        \draw [right] (16mm,-6mm) -- (16.1mm,-6mm);
        \draw [coil] (10mm,0) arc (-180:-90:6mm);
        \draw [mom] (22mm,0) arc (0:-90:6mm);
        \draw [right] (16mm,6mm) -- (16.1mm,6mm);
        \draw [coil] (10mm,0) arc (180:90:6mm);
        \draw [mom] (22mm,0) arc (0:90:6mm);
        \draw [aux,right] (22mm,0)  -- (27mm,0);
        \draw (27mm,0)  -- (31mm,0);
      \end{tikzpicture}%
    \end{gathered}
    \label{eq:hydro-sym-diagrams-2}\\
    \begin{gathered}
      \begin{tikzpicture}[thick]
        \draw [right] (0mm,0)  -- (5mm,0);
        \draw [aux] (5mm,0)  -- (10mm,0);
        \draw [right] (10mm,0) arc (-180:-90:6mm);
        \draw [mom] (10mm,0) arc (-180:0:6mm);
        \draw [right] (10mm,0) arc (180:90:6mm);
        \draw (22mm,0) arc (0:90:6mm);
        \draw [aux,right] (22mm,0)  -- (27mm,0);
        \draw (27mm,0)  -- (32mm,0);
      \end{tikzpicture}
  \end{gathered}
  \label{eq:hydro-sym-diagrams-3} \\
  \begin{gathered}
    \begin{tikzpicture}[thick]
      \draw [right] (0mm,0)  -- (5mm,0);
      \draw [aux] (5mm,0)  -- (10mm,0);
      \draw [right] (16mm,-6mm) -- (16.1mm,-6mm);
      \draw [mom] (10mm,0) arc (-180:0:6mm);
      \draw [right] (16mm,6mm) -- (16.1mm,6mm);
      \draw [mom] (10mm,0) arc (180:0:6mm);
      \draw [aux,right] (22mm,0)  -- (27mm,0);
      \draw (27mm,0)  -- (32mm,0);
    \end{tikzpicture}\qquad
    \begin{tikzpicture}[thick]
      \draw [right] (0mm,0)  -- (5mm,0);
      \draw [aux] (5mm,0)  -- (10mm,0);
      \draw [right] (16mm,-6mm) -- (16.1mm,-6mm);
      \draw [mom] (10mm,0) arc (-180:0:6mm);
      \draw [right] (16mm,6mm) -- (16.1mm,6mm);
      \draw [mom] (10mm,0) arc (180:90:6mm);
      \draw [coil] (22mm,0) arc (0:90:6mm);
      \draw [aux,right] (22mm,0)  -- (27mm,0);
      \draw (27mm,0)  -- (32mm,0);
    \end{tikzpicture}\qquad
    \begin{tikzpicture}[thick]
      \draw [right] (0mm,0)  -- (5mm,0);
      \draw [aux] (5mm,0)  -- (10mm,0);
      \draw [right] (16mm,-6mm) -- (16.1mm,-6mm);
      \draw [mom] (10mm,0) arc (-180:0:6mm);
      \draw [right] (16mm,6mm) -- (16.1mm,6mm);
      \draw [coil] (10mm,0) arc (180:90:6mm);
      \draw [mom] (22mm,0) arc (0:90:6mm);
      \draw [aux,right] (22mm,0)  -- (27mm,0);
      \draw (27mm,0)  -- (31mm,0);
    \end{tikzpicture}
  \end{gathered}
  \label{eq:hydro-sym-diagrams-4}
  \end{gather}
  \label{eq:hydro-sym-diagrams}%
\end{subequations}
The first diagrams in
\cref{eq:hydro-sym-diagrams-1,eq:hydro-sym-diagrams-2,eq:hydro-sym-diagrams-3}
contribute to leading order corrections in $\v[k]^2$, while the remaining
diagrams, along with those in \cref{eq:diff-sym-diagrams}, contribute at the
subleading order. We have computed these diagrams in Appendix
\ref{app:loop_details}, which yield a complicated expression given by
\begin{align}
  G^{\text{S},\text{1-loop}}_{nn}(\omega,\v[k])
  &=
    - \frac{T^2\v[k]^2}{(\omega+iD\v[k]^2)^2} \frac{\tilde\sigma}{\sigma}
    \Bigg[
    \frac{\chi}{w} \frac{\tilde\eta}{\eta} \mathcal{I}_0(p)
    - \frac{\tilde\sigma}{\sigma} \chi^2 \lambda^2 \mathcal{I}_1(p)
    + \frac{D\v[k]^2}{i\omega} \frac{\tilde\eta\chi_\epsilon}{w}
    \lb \lambda_\pi \mathcal{J}_1(p) - \psi\, \mathcal{J}_3(p)\rb 
    \nn\\
  &\qquad\qquad\qquad
    + \frac{n_0\tilde\eta}{w} \lb
    \lambda_\pi \mathcal{J}_1(p) + \lambda_\pi\mathcal{J}_2(p)
    - \psi \mathcal{J}_3(p) - \psi\mathcal{J}_4(p) \rb  \Bigg] \nn\\
  &\qquad
    - \frac{T^2\tilde\eta/w\,\v[k]^2}{\omega^2 + D^2\v[k]^4}
    \lb\frac{\tilde\eta}{\eta} - \frac{\tilde\sigma}{\sigma}\rb 
    \Bigg[
    \chi_\epsilon \frac{D\v[k]^2 }{i\omega}
    \lb \lambda_\pi \mathcal{J}_1(p) - \psi\, \mathcal{J}_3(p) \rb \nn\\
  &\qquad\qquad\qquad
    + n_0 \lb \lambda_\pi \mathcal{J}_1(p)
    + \lambda_\pi \mathcal{J}_2(p)
    - \psi\, \mathcal{J}_3(p) 
    - \psi \mathcal{J}_4(p) \rb \Bigg] \nn\\
  &\qquad
    - \frac{2iT^2\eta \tilde\sigma/\chi\,\v[k]^4}{
    (\omega+iD\v[k]^2)(\omega^2 + D^2\v[k]^4)} 
    \frac{\dow(\tilde\eta/\eta)}{\dow\mu}
    \lb\lambda_\pi \mathcal{J}_{2}(p) - \psi\mathcal{J}_{4}(p)\rb \nn\\
  &\qquad
    + \text{complex conjugate}.
\end{align}
When the KMS symmetry is respected, the result simplifies and the remaining
contributions match the relevant terms in the retarded correlation function, in accordance with
the fluctuation-dissipation theorem \eqref{fdt}. We have computed these terms in
the $\omega=0$ limit and verified that the answer is well-defined and still
non-analytic, even in the presence of KMS. For $d=3$, the results have been
explicitly presented in the introduction.


\section{Discussion}
\label{sec:discussion}

In this paper, we explored the spatial analyticity of hydrodynamic correlation
functions at loop-level, using the Schwinger-Keldysh EFT formalism that
incorporates nonlinear stochastic fluctuations consistent with the KMS
condition. We computed the complete one-loop correction to the retarded and
symmetric two-point functions of charge density, due to diffusive density and
transverse momentum fluctuations. The symmetric correlation function is found to
be generically non-analytic, owing to the absence of a characteristic mass scale
in the EFT. Despite this, we find that the retarded correlation function
non-trivially respects the spatial analyticity requirement expected in the
hydrostatic limit due to thermal screening. Notably, the analyticity behaviour
crucially depends on the dynamical KMS symmetry in the EFT, that leads to a
series of non-trivial cancellations of non-analytic terms in the hydrostatic
limit. Our final results are summarized in \cref{summary}.

We performed the majority of loop calculations in this paper in generic spatial
dimensions and away from the hydrostatic limit ($\omega\neq 0$). This analysis
can be recycled to study the phenomenon of long-time tails at finite-$\omega$
finite-$\v[k]$ in relativistic hydrodynamics in the presence of momentum
fluctuations, closely following a similar analysis for scalar diffusion model
in~\cite{Chen-Lin:2018kfl}. These results will appear in a companion
paper~\cite{other-paper}.

During the course of our analysis, we developed a novel EFT model for
``incompressible'' diffusive hydrodynamics that is consistent with the KMS
condition beyond the linearised fluctuations, accounting for the presence of
arbitrary non-linear interactions. This model can be understood as
systematically integrating out the ``higher-energy'' sound modes from the
hydrodynamic setup and allows us to study the effects of momentum fluctuations
in a controlled setting, which is expected to have far-reaching applications
beyond the context of the present work.

Although not the focus of this paper, in computing the symmetric correlation
functions, we have exhibited specific non-analyticities that are computable,
independent of the UV cutoff, and determined by thermodynamic parameters and
hydrodynamic transport coefficients. As such, symmetric correlators are
interesting quantities in themselves. One approach to computing transport
coefficients within the microscopic theory utilizes Euclidean correlators on a
finite size lattice. Although the physical content of Euclidean correlators most
directly maps to Minkowskian retarded correlators, considering the static limit
of the symmetric correlator may be helpful for extracting transport coefficients
via finite size scaling.

In terms of future directions, its important to note that we have limited our considerations in this paper to generic thermal systems that
are assumed to have a finite screened spatial correlation length. There are of course physical systems in which long-range correlations survive and
may lead to physical non-analyticities in retarded correlators. Examples include
conformal fixed points, and superfluids, which are known to require additional
degrees of freedom within the hydrostatic generating functional. It would be
interesting to explore how spatial non-analyticities re-emerge in such
theories. The analysis can also be extended in the context of non-relativistic
hydrodynamics using recently developed field theory
methods~\cite{Jain:2020vgc}.

Finally, we have argued that the KMS condition, associated with the existence
and properties of the thermal equilibrium state, is the critical ingredient in
ensuring the spatial analyticity of the retarded correlators in the hydrostatic
limit. While this is, of course, consistent with the assumptions underlying the
derivative expansion for the hydrostatic generating functional, it is natural to
ask whether it can be imposed in a more direct and local manner within the EFT
itself. Currently, the KMS condition is formulated as a non-local discrete
symmetry condition to be imposed on the Schwinger-Keldysh EFT, and it would be
interesting to understand if a more covariant formalism would allow for it to be
imposed directly in writing down the EFT Lagrangian. In particular, we have only
showed the analyticity of retarded correlators at one-loop level. It will be
interesting to explore whether such a result can be derived at the full
non-perturbative level in the EFT formalism of hydrodynamics.


\subsection*{Acknowledgments}

AS would like to acknowledge the hospitality of the Centre for Particle Theory at Durham
University, and the Department of Physics at McGill University, where parts of
this work were carried out. This work is supported in part by NSERC, Canada. Research at Perimeter Institute is supported in part by the Government of Canada through the Department of Innovation, Science and Economic Development Canada and by the Province of Ontario through the Ministry of Colleges and Universities.


\appendix

\section{The effective action for diffusive hydrodynamics}
\label{details}

In this appendix, we provide further details on the computation of the
perturbative effective action for relativistic diffusive hydrodynamics presented
in \cref{sec:hydroaction_expansion}, starting from the Schwinger-Keldysh
effective field theory.

Let us start with the effective action for full hydrodynamics in \cref{Lhydro},
with the constraints \cref{eq:incompressibility}. We need to expand the
effective action order-by-order in the dynamical fields $\delta\beta^\mu$,
$\delta\Lambda_\beta$, $X_a^\mu$, $\varphi_a$, and the background fields
$A_{rt}$, $A_{at}$. Segregating space and time derivatives in the effective
Lagrangian \cref{Lhydro}, we can express it as
\begin{align}
  \mathcal{L}
  &= (n_0 + \delta\hat n) A_{at}
    - \varphi_a \dow_t \delta\hat n
    - X^i_{a} \dow_t \hat\pi_i
    + J^i \dow_i \varphi_a
    + T^{ij} \dow_i X_{aj}
    + \delta\hat n\, X^i_{a} \dow_i A_{rt} \nn\\
  &\qquad
    + i T \tilde\sigma \dow_i \varphi_a \dow^i \varphi_a
    + i T \tilde \eta 
    \lb \dow^iX^j_{a} \dow_iX_{aj}
    + \dow^jX^i_{a} \dow_iX_{aj} \rb
    + \Psi_a \dow_i \frac{u^i}{T}
    + \Psi_r \dow_i X_a^i
    + \ldots,
    \label{eq:L-non-cov}
\end{align}
where we have only kept terms up to cubic order in dynamical and background
fields, and defined
\begin{align}
  \delta\hat n
  &\equiv J^t - n_0
    + 2i T \tilde\sigma u^i \dow_i \varphi_a \nn\\
  &= n -n_0
    + \half n_0  \v[u]^2
    - \sigma_0 u^i \lb \partial_i\mu - \dow_i A_{rt} \rb
    + 2i T_0 \tilde\sigma_0 u^i \dow_i \varphi_a
    + \cdots, \nn\\ 
  \hat\pi^i
  &\equiv T^{ti} + 2 i T \tilde\eta
    \lb u^k \dow_k X_{a}^i + u_k \dow^i X_a^k \rb \nn\\
  &= w\, u^i
    - \eta_0 \lb u^k \dow_k u^i + u_k \dow^i u^k \rb
    + 2 i T_0 \tilde\eta_0
    \lb u^k \dow_k X_{a}^i + u_k \dow^i X_a^k \rb
    + \cdots.
    \label{eq:actual_densities}
\end{align}
Here $w = \epsilon + p$ is the enthalpy density.  The subscript ``0'' denotes
that the corresponding quantity is evaluated at equilibrium. These are the true
non-classical conserved charge and momentum densities. We can further split
these into
\begin{align}
  \delta\hat n
  &= \delta n
    + \chi A_{rt}
    + \frac{\chi'}{\chi} \delta n A_{rt}
    + \half\chi' A_{rt}^2
    + \cdots, \nn\\
  \hat\pi^i
  &= \pi^i
    + \frac{\chi_\epsilon}{w_0} \pi^i A_{rt}
    + \cdots,
    \label{eq:dhn-dn}
\end{align}
with $\delta n$ and $\pi^i$ being the respective versions in
the absence of background fields,
\begin{align}
  \delta n
  &= \chi\delta\mu
    + \half\chi' \delta\mu^2
    + \half \chi_\epsilon \v[u]^2 
    - \sigma_0 u^i \partial_i\delta\mu
    + 2i T_0 \tilde\sigma_0 u^i \dow_i \varphi_a
    + \cdots, \nn\\
  \pi^i
  &= w_0 u^i
    + \chi_\epsilon u^i \delta\mu
    - \eta_0 \lb \half \dow^i \v[u]^2 + u^j \dow_j u^i \rb
    + 2 i T_0\tilde\eta_0 \lb u_j \dow^i X_{a}^j + u^j \dow_j X_{a}^i \rb
    + \cdots,
    \label{eq:dn-dmu}
\end{align}
where $\delta\mu = T_0\delta\Lambda_\beta$ and $u^i=T_0\delta\beta^i$, and we
have used
\begin{gather}
  \chi = \frac{\dow n}{\dow\mu}\Big|_0, \qquad
  \chi_{\epsilon} = n_0 + \frac{\dow \epsilon}{\dow\mu}\Big|_0
  = n_0 + \mu \frac{\dow n}{\dow\mu}\Big|_0
  + T\frac{\dow n}{\dow T}\Big|_0, \qquad
  \chi' = \frac{\dow^2 n}{\dow\mu^2}\Big|_0.
\end{gather}
Note that $\mu = \mu_0 + \delta\mu + A_{rt} + \mu_0/2\, \v[u]^2$. In the
following, we shall find it easier to work with $\delta n$, $\pi^i$ instead of
$\delta\mu$, $u^i$, because this eliminates time-derivative terms from the
interaction Lagrangian. We can find the inverted relations
\begin{align}
  \delta\mu
  &= \frac{1}{\chi} \delta n
    - \frac{\chi'}{2\chi^3} \delta n^2
    - \frac{\chi_\epsilon}{2\chi w_0^2} \v[\pi]^2 
    + \frac{\sigma_0}{\chi^2 w_0} \pi^i \partial_i\delta n
    - \frac{2i T_0\tilde\sigma_0}{\chi w_0} \pi^i \dow_i \varphi_a
    + \cdots, \nn\\ 
  u^i
  &= \frac{1}{w_0} \pi^i
    - \frac{\chi_\epsilon}{\chi w_0^2} \pi^i \delta n
    + \frac{\eta_0}{w_0^3} \lb \half\dow^i\v[\pi]^2 + \pi^j \dow_j \pi^i \rb
    - \frac{2 i T_0\tilde\eta_0}{w_0^2}
    \lb \pi_j \dow^iX_{a}^j + \pi^j \dow_jX_{a}^i \rb
    + \cdots.
\end{align}
The divergenceless condition in \cref{eq:incompressibility} implies that
\begin{equation}
  \dow_i\pi^i = \frac{\chi_\epsilon}{\chi w_0} \pi^i \dow_i\delta n
  + \ldots,
\label{divpi}
\end{equation}
where we have dropped the higher-derivative terms.

Using these expressions, we can go back and express $J^i$ and $T^{ij}$ also in
terms of the fluctuations $\delta n$ and $\pi^i$, and the background fields
$A_{rt}$ and $A_{at}$. We obtain
\begin{subequations}
  \begin{align}
    J^i
    &= n(T,\mu)\, u^i
      - \sigma(T,\mu) \delta^{ij} \lb T\partial_j
      \left(\frac{\mu}{T}\right) - \dow_j A_{rt} \rb
      - T_0\sigma_0 u^i \partial_t \left(\frac{\mu}{T}\right)
      + \ldots \nn\\
    &= \frac{n_0}{w_0} \pi^i
      - \frac{\sigma_0}{\chi} \partial^i \delta n
      - \lb \frac{n_0\chi_\epsilon}{\chi w_0^2} - \frac{1}{w_0} \rb \pi^i \delta n
      + \frac{n_0\eta_0}{w_0^3} \lb \half\dow^i\v[\pi]^2 + \pi^j \dow_j \pi^i \rb \nn\\
    &\qquad
      - \frac{1}{2\chi}\frac{\dow (\sigma/\chi)}{\dow \mu}\Big|_0
      \partial^i \delta n^2
      + \frac{\sigma_0 \chi_{\epsilon}}{2\chi w_0^2} \partial^i \v[\pi]^2
      - \frac{2 i T_0\tilde\eta_0n_0}{w_0^2}
      \lb \pi_j \dow^i X_{a}^j + \pi^j \dow_j X_{a}^i \rb \nn\\
    &\qquad
      + \frac{\chi}{w_0} A_{rt} \pi^i
      - \frac{1}{\chi}\frac{\dow \sigma}{\dow \mu}\Big|_0 A_{rt}
      \partial^i \delta n
      - \left\{\frac{\sigma_0}{\chi w_0} \pi^i \partial_\mu J^\mu \right\}
      + \ldots,
  \end{align}
  and
  \begin{align}
    T^{ij}
    &= w_0\, u^i u^j + p(T,\mu)\,\delta^{ij}
      - \eta(T,\mu) \lb\dow^i u^j + \dow^j u^i \rb
      - \eta_0 \lb u^i \dow_t u^j + u^j \dow_t u^i \rb
      + \ldots \nn\\
    &= - \frac{2\eta_0}{w_0} \dow^{(i} \pi^{j)} 
      + \frac{1}{w_0} \pi^i \pi^j 
      - \frac{2}{\chi w_0}\frac{\dow\eta}{\dow\mu}\Big|_0
      \delta n \dow^{(i}\pi^{j)}
      + \frac{2\eta_0\chi_\epsilon}{\chi w_0^2}  \delta n\, \dow^{(i} \pi^{j)}
      + \lb \frac{2\eta_0(n_0 + \chi_\epsilon)}{\chi w_0^2}  \rb \pi^{(i} \dow^{j)}\delta n \nn\\
    &\qquad
      - \frac{2}{w_0}\frac{\dow\eta}{\dow\mu}\Big|_0  A_{rt} \dow^{(i} \pi^{j)}
      - \left\{\frac{2\eta_0}{w_0^2}
      \pi^{(i} \lb \dow_\mu T^{j)\mu} - F^{j)\lambda}_r J_{\lambda} \rb \right\}
      + (\ldots) \delta^{ij} 
      + \ldots.
  \end{align}
  \label{eq:JiTijExpanded}%
\end{subequations}
We have not expanded the trace part of $T^{ij}$ as it couples to $\dow_i X_a^i$
in the effective action eq.\ \eqref{Lhydro} and does not contribute in the
diffusive limit eq.\ \eqref{eq:incompressibility}. We have expressed the terms
involving time-derivatives using the equations of motion in the expressions
above (denoted in braces), which can be removed by redefining $\varphi_a$ and
$X_{a}^i$ as
\begin{align}
  &\varphi_a \to \varphi_a
  - \frac{\sigma_0}{\chi w_0} \pi^i \dow_i\varphi_a\, , \qquad
  X_{a}^i \to X_{a}^i
  - \frac{\eta_0}{w_0^2}
  \lb \pi_k \dow^i X_{a}^k + \pi^k\dow_k X^i_a \rb.
\end{align}
Note that the imaginary part in $J^i$ is \emph{not} true statistical flux (which
will be obtained by varying the action), but merely one of the contributions to
the effective action.

The free, interaction, and background coupling part of the Lagrangian in
\cref{sec:hydroaction_expansion}, eqs.\ \eqref{freelang}, \eqref{Lhydroint}, and \eqref{sourlang}, follow directly from here by substituting for
$T^{ij}$, $J^i$, $\delta\hat n$, and $\hat\pi^i$ into \cref{eq:L-non-cov}.

\section{Loop calculations}
\label{app:loop_details}
In this appendix, we provide explicit details of the calculations involved in computing the loop
integrals presented in the main text.

One-loop corrections to the retarded and symmetric correlation functions of
charge density can be computed using the variational formulae in
\cref{eq:variational_formulas}. Starting with the retarded function, we first
need to compute the one-loop diagrams correcting the
$\langle\delta n\varphi_a\rangle$ propagator given in
\cref{eq:diff-ret-diagrams-prop,eq:hydro-ret-diagrams-prop}. Let us denote the amputated version, i.e. diagrams ignoring the external legs, by
$i\Gamma_{ar}(p)$. Next, we have diagrams involving a non-linear background
field coupling from \cref{eq:diff-ret-diagrams-back,eq:hydro-ret-diagrams-back},
whose amputated version is denoted by $i\Gamma'_{ar}(p)$. The full one-loop
retarded correlation function is given in terms of these as
\begin{equation}
  G^{\rm R,\text{1-loop}}_{nn}(p)
  = \frac{i\omega\chi \Gamma_{ar}(p)}{(i\omega-D\v[k]^2)^2}
  - \frac{\Gamma'_{ar}(p)}{i\omega-D\v[k]^2}.
  \label{eq:ret-expansion}
\end{equation}
For completeness, we note that the one-loop correction to the advanced
correlation function will be given by its complex conjugate,
\begin{equation}
  G^{\rm A,\text{1-loop}}_{nn}(p)
  = \frac{-i\omega\chi \Gamma^*_{ar}(p)}{(i\omega+D\v[k]^2)^2}
  + \frac{\Gamma'^*_{ar}(p)}{i\omega+D\v[k]^2}.
  \label{eq:adv-expansion}
\end{equation}
On the other hand, for the symmetric function, we need to compute the diagrams
in \cref{eq:hydro-sym-diagrams,eq:diff-sym-diagrams}. Note that the diagrams in
\cref{eq:diff-sym-diagrams-1,eq:hydro-sym-diagrams-1} are the same as those for the
retarded function in
\cref{eq:diff-ret-diagrams-prop,eq:hydro-ret-diagrams-prop}, except with
different external legs. Hence, their amputated version is still given by
$i\Gamma_{ar}(p)$. Furthermore, the diagrams in
\cref{eq:diff-sym-diagrams-2,eq:hydro-sym-diagrams-2} are merely their complex
conjugates, and hence their amputated version is given by
$-i\Gamma^*_{ar}(p)$. Therefore, the only diagrams that we need to compute for
the symmetric function are given in
\cref{eq:diff-sym-diagrams-3,eq:hydro-sym-diagrams-3,eq:hydro-sym-diagrams-4}. We
denote their amputated version by $\Gamma_{aa}$. In total, the one-loop
correction to the symmetric function is given by
\begin{equation}
  G^{\rm S,\text{1-loop}}_{nn}(p)
  = \frac{-\Gamma_{aa}(p)}{\omega^2+D^2\v[k]^4}
  - \frac{2T\tilde\sigma \v[k]^2\Gamma_{ar}(p)
  }{(i\omega-D\v[k]^2)(\omega^2+D^2\v[k]^4)}
  + \frac{2T\tilde\sigma \v[k]^2  \Gamma^*_{ar}(p)
  }{(i\omega+D\v[k]^2)(\omega^2+D^2\v[k]^4)}.
  \label{eq:sym-expansion}
\end{equation}
In the remainder of this appendix, we explicitly compute $\Gamma_{ar}(p)$,
$\Gamma'_{ar}(p)$, and $\Gamma_{aa}(p)$.

For clarity, we use the following notation: $F(p) = \omega + iD\v[k]^2$,
$G(p) = \omega+i\gamma_\eta\v[k]^2$,
$\bar k'^{ij} = \delta^{ij} - k'^i k'^j/\v[k]'^2$,
$\bar k''^{ij} = \delta^{ij} - k''^i k''^j/\v[k]''^2$, along with
$\omega'' = \omega - \omega'$ and $k''^i = k^i - k'^i$. We also denote the
frequency-momentum integrals compactly as
$\int_{p'} = \int \df\omega'\df^d\v[k]'/(2\pi)^{d+1}$, and just the momentum
integrals as $\int_{k'} = \int\df^d\v[k]'/(2\pi)^{d}$. Momentum integrals are
performed with a hard UV-cutoff at $|\v[k]'|=\Lambda$ and all the
cutoff-dependent terms are ignored. We make extensive use of the change of
variables $\omega'\to\omega''$, $\v[k]'\to\v[k]''$ to simplify the integrals,
which leads to identities such as
\begin{align}
  \int_{p'}\frac{\v[k]'^2}{F(p')F(p'')}
  &= \frac{1}{2}\int_{k'}
  \frac{\v[k]'^2+\v[k]''^2}{i\omega
    - D(\v[k]'^2+\v[k]''^2)}
    = \frac{i\omega}{2D}
    \int_{p'}\frac{1}{F(p')F(p'')}, \nn\\
  \int_{p'}\frac{\v[k]'\cdot\v[k]''}{F(p')F(p'')}
  &= \frac{1}{2}\int_{k'}
  \frac{\v[k]^2 - (\v[k]'^2+\v[k]''^2)}{i\omega
    - D(\v[k]'^2+\v[k]''^2)}
  = \frac{-iF(p)}{2D} \int_{p'}\frac{1}{F(p')F(p'')},
    \label{eq:manipulations}
\end{align}
up to cutoff-dependent terms.

\subsection{Computation of $\Gamma_{ar}$}

Let us start with the contributions from the two density fluctuation diagrams in
\cref{eq:diff-ret-diagrams-prop}. This is a generalization of the calculation
in~\cite{Jain:2020fsm} to account for $\tilde\sigma\neq\sigma$. We find
\begin{align}
  i\Gamma_{ar}(p)
  &\sim - 2T \tilde\sigma \lambda^2\v[k]^2\int_{p'}
    \frac{\v[k]'^2\v[k]''^2}{F(p')|F(p'')|^2}
    - iT\chi\lambda\tilde\lambda \v[k]^2 \int_{p'}
  \frac{\v[k]'\cdot \v[k]''}{F(p')F(p'')} \nn\\
  &= - iT\chi \lambda \v[k]^2 \int_{p'}
    \frac{\lambda\tilde\sigma/\sigma\, \v[k]'^2
    + \tilde\lambda\, \v[k]'\cdot \v[k]'' }{F(p')F(p'')} \nn\\
  &= - \frac{iT\chi \lambda}{D}
    \lb i\omega \lambda \frac{\tilde\sigma}{\sigma}
    - i F(p) \tilde\lambda \rb \mathcal{I}_1(p).
\end{align}
In the second step, we have split the tree ``$rr$'' propagator into a sum of
``$ra$'' and ``$ar$'' propagators according to \cref{eq:diff-propagators}, and
noted that the ``$ar$'' piece does not contribute due to a purely retarded loop.
The integral $\mathcal{I}_1(p)$ is defined in \cref{eq:I1}. Next, we have
contributions coming from the single leading-order ``mixed'' fluctuation diagram
in \cref{eq:hydro-ret-diagrams-prop}
\begin{align}
  i\Gamma_{ar}(p)
  &\sim \frac{2T\tilde\eta}{w^2}\int_{p'}
    \frac{\v[k]'^2\, \bar k'^{ij} k_i k''_j}{|G(p')|^2F(p'')}
    = \frac{iT}{w} \frac{\tilde\eta}{\eta} \int_{p'}
    \frac{\v[k]^2 - (\v[k]\cdot \v[k]')^2/\v[k]'^2}{G(p')F(p'')}
    = \frac{iT\v[k]^2}{w} \frac{\tilde\eta}{\eta} \mathcal{I}_0(p).
\end{align}
The integral $\mathcal{I}_0(p)$ is defined in \cref{eq:I0}. Finally, we have two
momentum-fluctuation diagrams in \cref{eq:hydro-ret-diagrams-prop} leading to
\begin{align}
  i\Gamma_{ar}(p)
  &\sim 2T\tilde\eta  \int_{p'}
    \frac{\v[k]''^2
    \lb \lambda_\pi \v[k]^2 \delta_{ij} - 2\psi k_i k_j \rb
   \bar k'^{ik} \bar k''^{jl}
  \lb \gamma \delta_{kl} \v[k]'^2
  + \theta (\delta_{kl} \v[k]'\cdot\v[k]'' + k'_l k''_k) \rb}
    {G(p')|G(p'')|^2} \nn\\
  &\qquad
    - iTw\tilde\theta \int_{p'}
  \frac{\lb \lambda_\pi \v[k]^2 \delta_{ij} - 2\psi k_i k_j \rb
  \bar k'^{ik} \bar k''^{jl}
   (\delta_{kl}\v[k]'\cdot \v[k]'' + k'_l k''_k)}
    {G(p')G(p'')} \nn\\
  &= iTw \int_{p'}
    \frac{\lb \lambda_\pi \v[k]^2 \delta_{ij} - 2\psi k_i k_j \rb
   \bar k'^{ik} \bar k''^{jl}
  \lb \frac{\tilde\eta}{\eta} \gamma \delta_{kl} \v[k]'^2
    + \lb \frac{\tilde\eta}{\eta} \theta - \tilde\theta \rb
    (\delta_{kl} \v[k]'\cdot\v[k]'' + k_k k_l) \rb}
    {G(p')G(p'')} \nn\\
  &= 
    iTw \v[k]^2 \lb
    \frac{\tilde\eta}{\eta}\lambda_\pi \gamma\, \mathcal{J}_1(p)
    + \lambda_\pi \lb \frac{\tilde\eta}{\eta} \theta - \tilde\theta \rb
    \mathcal{J}_2(p)
    - \frac{\tilde\eta}{\eta}\psi \gamma\, \mathcal{J}_3(p)
    - \psi \lb \frac{\tilde\eta}{\eta} \theta - \tilde\theta \rb
    \mathcal{J}_4(p)
    \rb.
    \label{eq:Gamma-ar}
\end{align}
The integrals $\mathcal{J}_{1,2,3,4}(p)$ are defined as
\begin{align}
  \mathcal{J}_1(p)
  &= \int_{p'}
    \frac{\v[k]''^2\bar k'^{ik}\bar k''^{jl} \delta_{ij}\delta_{kl}}
    {G(p')G(p'')}
    = \int_{p'} \frac{\v[k]''^2(d-2) + (\v[k]'\cdot \v[k]'')^2/\v[k]'^2}
    {G(p')G(p'')}, \nn\\[1ex]
  \mathcal{J}_2(p)
  &= \int_{p'}
    \frac{\bar k'^{ik}\bar k''^{jl}\delta_{ij}
    \lb \v[k]'\cdot\v[k]''\delta_{kl} + k_k k_l\rb}
    {G(p')G(p'')}
    = \int_{p'}
    \frac{(\v[k]'\cdot \v[k]'')
    ((d-3) + 2(\v[k]'\cdot \v[k]'')^2/(\v[k]'^2\v[k]''^2))}
    {G(p')G(p'')}, \nn\\[1ex]
  \mathcal{J}_3(p)
  &= \frac{2}{\v[k]^2}\int_{p'}
    \frac{\v[k]''^2 \bar k'^{ik}\bar k''^{jl}\delta_{ij} k_k k_l}
    {G(p')G(p'')}
    = \frac{2}{\v[k]^2} \int_{p'}
    \frac{(\v[k]'\cdot\v[k]'')(-\v[k]''^2 + (\v[k]'\cdot\v[k]'')^2/\v[k]'^2)}
    {G(p')G(p'')}, \nn\\[1ex]
  \mathcal{J}_4(p)
  &= \frac{2}{\v[k]^2} \int_{p'}
    \frac{\bar k'^{ik}\bar k''^{jl} k_i k_j
    \lb \v[k]'\cdot\v[k]''\delta_{kl} + k_k k_l \rb }
    {G(p')G(p'')}
   = \frac{2}{\v[k]^2} \int_{p'}
    \frac{\v[k]'^2\v[k]''^2 - 3(\v[k]'\cdot \v[k]'')^2
    + 2(\v[k]'\cdot \v[k]'')^4/(\v[k]'^2\v[k]''^2)}{G(p')G(p'')}.
\end{align}
Employing identities similar to those in \cref{eq:manipulations}, these four
integrals can be related to just two integrals $\mathcal{I}_2(p)$,
$\mathcal{I}_3(p)$, as given in \cref{eq:Jexpansion}.

\subsection{Computation of $\Gamma'_{ar}$}

For the two density fluctuation background coupling diagrams in
\cref{eq:diff-ret-diagrams-back}, we find
\begin{align}
  i \Gamma'_{ar}(p)
  &\sim
    \frac{2T\tilde\sigma \lambda}{\chi} \v[k]^2
    \int_{p'} \frac{\v[k]''^2(\sigma' \v[k]'\cdot\v[k]'' +
    \chi' i\omega')}{F(p')|F(p'')|^2 }
    - iT \chi^2 \lambda\tilde\lambda \v[k]^2
    \int_{p'} \frac{\v[k]'\cdot\v[k]''}{F(p')F(p'')} \nn\\
  &= iT\lambda \v[k]^2 
    \int_{p'}
    \frac{\frac{\tilde\sigma}{\sigma} 
    \chi' i\omega' 
    + \lb \frac{\tilde\sigma}{\sigma} \frac{\dow\sigma}{\dow\mu}
    - \frac{\dow\tilde\sigma}{\dow\mu} \rb
    \v[k]'\cdot\v[k]'' }{F(p')F(p'')} \nn\\
  &= \frac{iT\lambda }{D}
    \lb \frac{\tilde\sigma}{\sigma} 
    \chi'D i\omega
    + i F(p)\sigma \frac{\dow(\tilde\sigma/\sigma)}{\dow\mu}
    \rb \mathcal{I}_1(p).
\end{align}
The two leading-order mixed fluctuation diagrams in
\cref{eq:hydro-ret-diagrams-back} lead to
\begin{align}
  i\Gamma'_{ar}(p)
  &\sim \frac{2T\tilde\eta \chi}{w^2}\int_{p'}
    \frac{\v[k]'^2 \bar k'^{ij} k''_ik''_j}{|G(p')|^2F(p'')}
    - \frac{2T\tilde\sigma }{w} \int_{p'}
    \frac{\v[k]''^2 \bar k'^{ij} k_i k''_j}{G(p')|F(p'')|^2} \nn\\
  &= \frac{iT \chi}{w}
    \lb \frac{\tilde\eta}{\eta} - \frac{\tilde\sigma}{\sigma} \rb \int_{p'}
    \frac{\v[k]^2 - (\v[k]\cdot\v[k]')/\v[k]'^2}{G(p')F(p'')} \nn\\
  &= \frac{iT \chi \v[k]^2}{w}
    \lb \frac{\tilde\eta}{\eta} - \frac{\tilde\sigma}{\sigma} \rb \mathcal{I}_{0}(p).
\end{align}
Finally, the remaining two momentum fluctuation diagrams in
\cref{eq:hydro-ret-diagrams-back} evaluate to
\begin{align}
  i\Gamma'_{ar}(p)
  &= \frac{2T\tilde\eta}{w} \int_{p'}
    \frac{\v[k]''^2 \lb \lambda_\pi \v[k]^2 \delta_{ij} - 2\psi k_i k_j \rb
    \bar k'^{ik} \bar k''^{jl}
    \lb \chi_\epsilon i\omega' \delta_{kl}
    + \frac{\dow \eta}{\dow\mu} (\delta_{kl}\v[k]'\cdot\v[k]'' + k'_l k''_k) \rb}
    {G(p')|G(p'')|^2} \nn\\
  &\qquad
    - T\chi w\tilde\theta 
    i\int \frac{\df\omega'\df^dk'}{(2\pi)^{d+1}}
    \frac{\lb \lambda_\pi \v[k]^2 \delta_{ij} - 2\psi k_i k_j \rb
    k'^{ik} k''^{jl}
    (\delta_{kl}\v[k]'\cdot\v[k]'' + k'_l k''_k)}
    {G(p')G(p'')} \nn\\
  &= iT \int_{p'}
    \frac{\lb \lambda_\pi \v[k]^2 \delta_{ij} - 2\psi k_i k_j \rb
    \bar k'^{ik} \bar k''^{jl}
    \lb \frac{\tilde\eta}{\eta} \chi_\epsilon i\omega'\delta_{kl}
    + \lb\frac{\tilde\eta}{\eta}\frac{\dow\eta}{\dow\mu}
    - \frac{\dow\tilde\eta}{\dow\mu} \rb
    (\delta_{kl}\v[k]'\cdot\v[k]'' + k_l k_k) \rb}
    {G(p')G(p'')} \nn\\
  &= iT \lambda_\pi\v[k]^2 \lb \frac{\tilde\eta}{w} \chi_\epsilon
    \lb \mathcal{J}_1(p) + \half \mathcal{I}_3(0,\v[k]) \rb
    + \eta \frac{\dow(\tilde\eta/\eta)}{\dow\mu} \mathcal{J}_2(p) \rb \nn\\
  &\qquad
    - iT \psi \v[k]^2 \lb
    \frac{\tilde\eta}{w} \chi_\epsilon
    \lb \mathcal{J}_3(p) + \half \mathcal{I}_3(0,\v[k]) \rb
    + \eta \frac{\dow(\tilde\eta/\eta)}{\dow\mu} \mathcal{J}_4(p) \rb.
\end{align}
These expressions can be combined with the ones for $\Gamma_{ar}$ to obtain the
retarded correlation function following \cref{eq:ret-expansion}.

\subsection{Computation of $\Gamma_{aa}$}

We start with the three density fluctuation diagrams in
\cref{eq:diff-sym-diagrams-3}, resulting in
\begin{align}
  \Gamma_{aa}(p)
  &\sim - 2T^2\tilde\sigma^2\lambda^2\v[k]^4 \int_{p'}
    \frac{\v[k]'^2 \v[k]''^2}{|F(p')|^2|F(p'')|^2}
    + \lb 4iT^2\chi\lambda\tilde\lambda \tilde\sigma \int_{p'}
    \frac{\v[k]''^2\, \v[k]^2 (\v[k]\cdot\v[k]')}{F(p')|F(p'')|^2}
    + \text{c.c.} \rb  \nn\\
  &= T^2\chi^2\lambda\v[k]^2 \frac{\tilde\sigma}{\sigma}
    \int_{p'}
    \frac{\half\lambda \tilde\sigma/\sigma\,\v[k]^2
    - 2\tilde\lambda (\v[k]\cdot\v[k]') }{F(p')F(p'')}
    + \text{c.c.} \nn\\
  &=
    T^2\chi^2\lambda\v[k]^2 \frac{\tilde\sigma}{\sigma}
    \lb \frac{\tilde\sigma}{\sigma} \lambda
    - 2\tilde\lambda \rb \mathcal{I}_1(p)
    + \text{c.c.}
\end{align}
Here ``c.c.'' denotes complex conjugate. Note that the last two diagrams in
\cref{eq:diff-sym-diagrams-3} are complex conjugates of each other. Similarly,
we can compute the leading order mixed fluctuation diagram in
\cref{eq:hydro-sym-diagrams-3},
\begin{align}
  \Gamma_{aa}(p)
  &\sim - \frac{4T^2\tilde\sigma\tilde\eta}{w^2} \int_{p'}
    \frac{\v[k]'^2\v[k]''^2\, \bar k'^{ij}k_ik_j}{|G(p')|^2|F(p'')|^2} \nn\\
  &= \frac{T^2\chi}{w}
    \frac{\tilde\sigma}{\sigma} \frac{\tilde\eta}{\eta} \int_{p'}
    \frac{\v[k]^2 - (\v[k]\cdot\v[k]')/\v[k]'^2}{G(p')F(p'')}
    + \text{c.c.} \nn\\
  &= \frac{T^2\chi\v[k]^2}{w}
    \frac{\tilde\sigma}{\sigma} \frac{\tilde\eta}{\eta} \mathcal{I}_0(p)
    + \text{c.c.}
\end{align}
This leaves us with three momentum fluctuation diagrams in
\cref{eq:hydro-sym-diagrams-4}, which can be computed to obtain
\begin{align}
  \Gamma_{aa}(p)
  &\sim - 2 T^2\tilde\eta^2  \int_{p'}
    \frac{\v[k]'^2\v[k]''^2 \lb \lambda_\pi \delta_{ij}\v[k]^2 - 2\psi k_i k_j \rb
    \bar k'^{ik}\bar k''^{jl}
    \lb \lambda_\pi \delta_{kl}\v[k]^2 - 2\psi k_k k_l \rb
    }{|G(p')|^2|G(p'')|^2} \nn\\
  &\qquad
    - \lb 2i T^2 w \tilde\psi \tilde\eta \int_{p'}
    \frac{\v[k]''^2 (\lambda_\pi \delta_{ij} \v[k]^2 - 2\psi k_i k_j)
    \bar k'^{ik}\bar k''^{jl}
    (\delta_{kl} \v[k]\cdot\v[k]' + k_k k'_l)}{G(p')|G(p'')|^2}
    + \text{c.c.} \rb \nn\\
  &= T^2w^2 \frac{\tilde\eta}{\eta} \int_{p'}
    \frac{\lb \lambda_\pi \delta_{ij}\v[k]^2 - 2\psi k_i k_j \rb
    \bar k'^{ik}\bar k''^{jl}
    \lb \half \lambda_\pi \frac{\tilde\eta}{\eta} \delta_{kl}\v[k]^2
    + \frac{2n_0\tilde\eta}{w^3} \delta_{kl} \v[k]\cdot\v[k]'
    + \frac{n_0\tilde\eta}{w^3} k_k k_l \rb
    }{G(p')G(p'')}
    + \text{c.c.} \nn\\
  &= \frac{T^2\tilde\eta\v[k]^2}{w}  \frac{\tilde\eta}{\eta}
    \lb
    \lb \lambda_\pi \mathcal{J}_1(p) - \psi\, \mathcal{J}_3(p) \rb
    \lb n_0 - \chi_\epsilon \frac{D\v[k]^2 }{i\omega} \rb
    + n_0 \lambda_\pi \mathcal{J}_2(p) - n_0 \psi \mathcal{J}_4(p) \rb
    + \text{c.c.}
\end{align}
This result can be used in conjunction with $\Gamma_{ar}$ to obtain the symmetric
correlation function using \cref{eq:sym-expansion}.


\bibliographystyle{JHEP}
\bibliography{refs}

\end{document}